\documentclass{article}

\usepackage{enumerate}
\usepackage{amsfonts}
\usepackage{amsmath}
\usepackage{comment}
\usepackage{textcomp}
\usepackage{amsthm, tikz}
\usepackage{amssymb}
\usepackage{color}
\usepackage{graphicx}
\usepackage{bbm}
\usepackage[matrix,arrow,curve]{xy}
\usepackage{array}
\usepackage{mathtools}

\def\be{\begin{eqnarray}}
\def\ee{\end{eqnarray}}
\def\nn{\nonumber}

\usepackage{tikz}
\usepackage{braids}
\def\l[{\phantom.[}

\hoffset= - 1.0in         % switch off for draft style
\begin{document}

\hfill ITEP/TH-11/16

\hfill IITP/TH-09/16

\bigskip

\centerline{\Large{Differential expansion and rectangular HOMFLY
for the figure eight knot
%$4_1$
}}

\bigskip

\bigskip

\centerline{\bf  A.Morozov }

\bigskip

{\footnotesize
\centerline{{\it
ITEP, Moscow 117218, Russia}}

\centerline{{\it
Institute for Information Transmission Problems, Moscow 127994, Russia
}}

\centerline{{\it
National Research Nuclear University MEPhI, Moscow 115409, Russia
}}
}

\bigskip

\bigskip

\centerline{ABSTRACT}

\bigskip

{\footnotesize
Differential expansion (DE) for a Wilson loop average in representation $R$
is built to respect degenerations of representations for small groups.
At the same time it behaves nicely under some changes of the loop,
e.g. of some knots in the case of $3d$ Chern-Simons theory.
Especially simple is the relation between the DE
for the trefoil $3_1$ and for the figure eight knot $4_1$.
Since arbitrary colored HOMFLY for the trefoil are known from
the Rosso-Jones formula, it is therefore enough to find their DE
in order to make a conjecture  for the figure eight.
We fulfil this program for all rectangular representation $R=[r^s]$,
i.e. make a plausible conjecture for the rectangularly colored
HOMFLY of the figure eight knot, which generalizes the old result
for totally symmetric and antisymmetric representations.
}

\bigskip

\bigskip

\section{Introduction}

Chern-Simons (CS) theory \cite{CS} lies at the boundary between two very different
worlds -- of Yang-Mills  and of topological theories.
Because of this it serves as a bridge, allowing transfer of ideas and methods
between the two fields.
As topological theory, CS is exactly solvable -- in the sense that any particular
quantity (correlator) can be calculated, if one applies enough skill and effort.
Even if there are some indications of chaos \cite{KoMand}, they are well under control,
in the spirit of \cite{DMmand}.
It is not immediately like this in truly dynamical Yang-Mills theory, where quantities with
regular behavior at all energy scales and/or all time moments are rather rare
and difficult to identify.
At the same time, observables in CS theory
(known as knot polynomials \cite{knotpols}-\cite{knotebook})
depend on the same parameters -- group and representations -- as in generic
Yang-Mills theory, and this provides a possibility to study these
dependencies, separated from obscure space-time and energy-momentum properties.
From this point of view of special interest are the aspects of knot-polynomial
calculus, which rely not  so much on topological invariance,
but rather on the group- and representation theory properties,
common for all Yang-Mills theories.
Such are, for example, the quasiclassical and genus expansions
(known as Vassiliev and Hurwitz expansions in knot theory) and the AMM/EO {\it topological
recursion} \cite{AMM/EO} in the latter case (this time "topological" refers not to topological
theory, but to the structure of Feynman diagrams and/or spectral surfaces --
which are also characteristics of theories with real dynamics).
In fact, these two do not exhaust the interesting structures in Yang-Mills theories  --
among less known the most intriguing is the {\it differential expansion} (DE).
The word "differential" here refers to technical(?) connection to Khovanov's
differential in the presentation of \cite{DGR}, which have a lot to do with
the topological aspects of knot theory.
However, the DE itself is rather a pure representation-theory property,
reflecting the fact that different representations can occasionally coincide
for small groups.
Despite very simple, this fact provides unexpectedly much information about
the observables (knot polynomials).

The study of DE actually began in \cite{IMMMfe}, which was a part of a broad renewed
attack on the problem of knot polynomials and Racah matrices \cite{RT}-\cite{mmms3}.
In \cite{IMMMfe} DE was used to conjecture
a general expression for HOMFLY and superpolynomials of the very simple
figure eight knot $4_1$ in all symmetric and antisymmetric representations.
Later these formulas were extended to many more knots \cite{MMMevo,arthdiff,arthAENV}
and also used to obtain the {\it exclusive} Racah matrices \cite{gmmms,nrz} --
which, after conjectured, provide a systhematic approach to calculations for
all arborescent knots \cite{mmmrv}-\cite{mmms3}.
Despite this tremendous success, the DE method is thought to be too difficult
and does not attract much attention
-- except for serious developments in \cite{ano21,twist,Konodef,MMMS21}.
It is the goal of the present paper to once again demonstrate its abilities.
We do this by conjecturing an extension of  \cite{IMMMfe} for $4_1$
from symmetric and antisymmetric to arbitrary rectangular representations
(labeled by rectangular Young diagrams $R=[r^s]$ with $r$ columns and $s$ rows).
This is tedious, but surprisingly straightforward.
Among next challenges the first one is generalization from $4_1$ to other twist
and, further, double-braid knots of \cite{MMMevo}, because then one will be able to
apply  the double-evolution technique from \cite{mmms3} to deduce exclusive
Racah matrices $\bar S$ and $S$ -- and then calculate rectangularly-colored
HOMFLY for arbitrary arborescent knots.
This, however, is beyond the scope of the present text, which is concentrated on $4_1$.

In sec.\ref{gen} we provide a brief review \cite{Konodef} of the properties of
differential expansions and their enhancement for defect-zero knots, like
trefoil, figure eight, twist and double- braid families.
In remaining sections we outline step by step the technique to build the DE for the known
(from Rosso-Jones formula \cite{RJ}) rectangular HOMFLY of the trefoil --
known are the polynomials, but their additional structure, DE, needs to be revealed,
and this is the most difficult part of the story.
However, once revealed, it is very easily deformed from $3_1$ to $4_1$
(and, hopefully, also for other twist and double-braid knots).
This deformation provides the main result of the present paper --
the answer for $H^{4_1}_{[r^s]}$.
In this case the 3-graded super- and hyperpolynomials, as well as the 4-graded version
of the latter \cite{GGS},
are  provided by the changes of variables \cite{IMMMfe,MMMevo,arthdiff}.
We end in sec.\ref{conc} with a short conclusion.

\section{Generalities
\label{gen}}

Differential expansion (DE) from \cite{IMMMfe} for normalized knot polynomials
of the figure-eight knot ${\cal K}=4_1$ in any symmetric representation $R=[r]$ is:
\be
H^{4_1}_{[r]} = 1 + [r]\{Aq^{r+1}\}\{A/q\} + \frac{[r][r-1]}{[2]}\{Aq^{r+2}\}\{Aq^{r+1}\}\{A\}\{A/q\}
+\ldots = \nn \\
= \sum_{k=0}^r \frac{[r]!}{[k]![r-k]!}\prod_{i=0}^{k-1}\{Aq^{r+i}\}\{Aq^{i-1}\}
\label{desymm41}
\ee
where  $\{x\} = x-x^{-1}$, quantum numbers are defined as $[n]=\frac{\{q^n\}}{\{q\}}$
and quantities $D_n=\{Aq^n\}$ are often called {\it differentials}
(the name comes from Khovanov calculus {\it a la}
\cite{DGR}, where multiplication by $A^2q^{2n}$ is the main operation, responsible for
building the Khovanov-Rozansky complexes and their colored generalizations).
It follows that for antisymmetric representations $R=[1^r]$
\be
H^{4_1}_{[1^r]}(A,q) = H^{4_1}_{[r]}\left(A,\frac{1}{q}\right)
%= 1 + [r]\{Aq^{r+1}\}\{A/q\} + \frac{[r][r-1]}{[2]}\{Aq^{r+2}\}\{Aq^{r+1}\}\{A\}\{A/q\} +\ldots
= \sum_{k=0}^r \frac{[r]!}{[k]![r-k]!}\prod_{i=0}^{k-1}\{Aq^{1-i}\}\{Aq^{-r-i}\}
\label{deasymm41}
\ee
DE was further generalized to all twist knots in \cite{MMMevo}
and to all knots and even links in \cite{arthdiff,arthAENV} and finally in \cite{Konodef}.
In general
\be
H^{{\cal K}^{(d)}}_{[r]}\!\!(A,q) =
\sum_{k=0}^r \frac{[r]!}{[k]![r-k]!}\,G_k^{{\cal K}^{(d)}}\!\!(A,q)
\prod_{i=0}^{k-1}\{Aq^{r+i}\}\prod_{i=0}^{k-1-d}\{Aq^{i-1}\}
\label{desymmKd}
\ee
where parameter $d$ is an important characteristic of the knot ${\cal K}$,
called the {\it defect} of DE -- from (\ref{desymm41}) we see that defect is zero for ${\cal K}=4_1$.
As found in \cite{Konodef}, $d+1$ is actually the degree in $q^{\pm 2}$
of the fundamental Alexander polynomial
$Al_{_\Box}(q) = H_{_\Box}(A=1,q) = Al_{_\Box}(q^{-1})$  (for the figure eight knot
$Al_{_\Box}^{4_1}(q) = 3-q^2-q^{-2}$).
Accordingly the defect can be as small as $d=-1$ -- this happens for the first time
in the Rolfsen table \cite{katlas} for a pair of Kinoshita-Terasaks and Conway
11-crossing mutant knots with unit Alexander.
For numerous examples of DE in symmetric representations
see \cite{mmms1,mmms2,mmms3} and references therein.
The remarkable fact is  that $G_k$ in (\ref{desymmKd}) do not depend on $r$.

For attempts to preserve this property in generalization from symmetric to other representations
see \cite{ano21,MMMS21}.
From factorization property of special polynomials \cite{DMMSS,anton},
\be
H_R(A,q=1) =  H_{_\Box}(A,q=1)\Big)^{|R|} \ \ \ \ \forall \ R
\ee
it follows that corrections to (\ref{desymmKd}) for non-symmetric representations should
be proportional to $\{q\}$.
For single-hook representations $R=[r,1^s]$ there is a {\it dual} property for Alexander
polynomials \cite{DMMSS}
\be
H_R(A=1,q) = \Big(H_{_\Box}\left(A=1,q^{|R|}\right) \ \ \ \ \ {\rm for} \ \ \ R=[r,1^s]
\ee
corrections for these representations should should also be proportional to $\{A\}$.
Note also that this latter property implies that
whenever $G_1$ depends on $q$, i.e. defect is greater than zero (Alexander power is greater than one),
the higher $G_k$ can not vanish at $A=1$ -- otherwise it is impossible to preserve
$Al_{[r]}(q)=Al_{_\Box}(q^r)$.

For our purposes in the present paper important are the following properties of the
differential expansion:

$\bullet$ DE represents knot polynomials
as polynomials of degree $|R|+1$ in the differentials $D_n$.

$\bullet$ Coefficients in these polynomials are functions of $q$ and $A$, so it is not quite
easy to give a formal definition of the expansion.

$\bullet$
DE is also a version of Vassiliev expansion in $h$ for $q=1+h$ and $A=(1+h)^N$ --
with this definition HOMFLY modulo a framing factor are polynomials, not series, in $h$ --
still again $D_n \sim h$, but the coefficients also depend on $h$.

$\bullet$ The shape of DE is partly dictated by the  fact, that {\bf knot polynomial
depends on representation, i.e. when representations coincide, the same is true
about knot polynomials}.

In this paper we mostly elaborate on the boldfaced statement in the list.
This simple fact actually stands behind the "surprising" success of differential expansion method
for symmetric representations $R$ -- and it remains quite powerful for arbitrary rectangular $R$.

\section{Restrictions on differential expansion from group theory}

We shall use the combination of three facts:

$\bullet$ For $A=q^N$ $H_R(A,q)$ depends on representation $R$ of $SU(N)$

$\bullet$ For $A=q^{-N}$ the transposed $H_R(A,q^{-1})=H_{R^{tr}}(A,q)$
depends on representation $R^{tr}$ of $SU(N)$

$\bullet$ For particular $SU(N)$ there are special relations between
conjugate representations with $l$ and $N-l$ lines

\noindent
For rectangular Young diagrams $[r^s]$
representation $[r^s]\ \stackrel{SU(s)}{\cong}\ \emptyset$ is
equivalent to a singlet $\emptyset$ for $SU(s)$
and likewise $\widetilde{[r^s]} = [s^r] \ \stackrel{SU(r)}{\cong} \ \emptyset$,
where we use tilde to denote transposition of Young diagrams.
Then the above three facts imply that
\be
H_{[r^s]}(A=q^s,q) = 1
\ \ \ \ \ \ \ \ \ \ \
{\rm and}
\ \ \ \ \ \ \ \ \ \ \
H_{[r^s]}(A=q^{-r},q) = 1
\label{preHrs1}
\ee
More generally
\be
[r^s] \ \stackrel{SU(N)}{\cong}\  [r^{N-s}]
\ \ \ \ \ \ \ \ \ \
{\rm and}
\ \ \ \ \ \ \ \ \ \
\widetilde{[r^s]}=[s^r] \ \stackrel{SU(N)}{\cong}\  [s^{N-r}] =\widetilde{[(N-r)^{s}]}
\label{preHrsN}
\ee
what imposes severe constraints on the next terms of the differential expansion.
The only word of caution is that in above relations $N$ should not be taken smaller than $r$ or $s$ --
trivialization of representations with the number of lines $l_R>N$
implies {\it nothing} for {\it normalized} knot polynomials -- what vanishes in these
cases are {\it dimensions} ${\rm dim}_R$, while {\it normalized} polynomials
stay non-trivial.

As a warm-up, let us look at symmetric and antisymmetric representations $[r]$ and
$[1^r]=\widetilde{[r]}$.
Since for $U(1)$ all $[r] \ \stackrel{U(1)}{\cong} \ \emptyset$ are trivial, we conclude that
$H_{[r]}(A=q)=1$, i.e.
\be
H_{[r]}-1 \sim \{A/q\}
\label{HsymmU1}
\ee
Note, that "transposed" statement $H_{[1]}-1\sim \{Aq\}$ for $A=q^{-1}$ is true only for
$r=1$ -- because of the above-mentioned restriction $l\leq N$ on the number $l_{[1^r]}=r$ of lines of
transposed diagram $\widetilde{[r]} = [1^r]$.
Instead, from $[1^r]\ \cong{SU(r)}{\cong}\ \emptyset$ we get:
\be
H_{[1^r]} -1 \sim \{A/q^r\} \ \ \Longleftrightarrow \ \
H_{[r]} - 1 \sim \{Aq^r\}
\ee
and, together  with (\ref{HsymmU1}),
\be
H_{[r]}^{\cal K} -1 \sim \{Aq^r\}\{A/q\}
\label{Hrfirst}
\ee
This is in obvious accordance with (\ref{desymm41})
and, as we see, this is true for {\it arbitrary} knots ${\cal K}$:

Restrictions on the higher terms of the differential expansion come from
\be
[1^r]\ \cong{SU(N)}{\cong}\ [1^{N-r}]
\label{rcongNmr}
\ee
with $N>r$.
%Indeed, if we accept that
%
%$\bullet$  differential expansion is a polynomial of degree $|R|+1$
%in the differentials
%(the main motivation for this comes from the study of {\it special} polynomials
%at $q=1$),
%
%\noindent
%then (\ref{rcongNmr}) relates polynomials of different degrees
%modulo $D_N=\{Aq^N\}$.
For example, for $N=3$
\be
H_{[11]}-H_{[1]} \sim \{A/q^3\} \ \ \Longleftrightarrow \ \
H_{[2]} - H_{[1]} \sim \{Aq^3\} \ \ \stackrel{(\ref{HsymmU1})}{\Longrightarrow}
H_{[2]} - H_{[1]} \sim \{Aq^3\} \{\{A/q\}
\ee
Denoting the proportionality coefficients by $G_1^{\cal K}(A,q)$ and $g_2^{\cal K}(A,q)$
we get:
\be
H_{[1]} = 1+G_1\cdot \{Aq\}\{A/q\}, \nn \\
H_{[2]} = H_{[1]} + g_2\cdot \{Aq^3\}\{A/q\}
\ee
and this should be now combined with (\ref{Hrfirst}):
\be
H_{[2]} = 1 + \Big(G_1\{Aq\}+g_2\{Aq^3\}\Big)\{A/q\} = 1 + \tilde g_2 \{Aq^2\}\{A/q\}
\ee
Since $\{Aq^3\}+\{Aq\} = [2]\{Aq^2\}$, it follows that $g_2 = G_1 + G_2\{Aq^2\}$,
$\tilde g_2 = [2]G_1 + G_2\{Aq^3\}$ for some $G_2$
and
\be
H_{[1]} = 1+G_1\cdot \{Aq\}\{A/q\}, \nn \\
H_{[2]} = 1 + [2]G_1\cdot\{Aq^2\}\{A/q\} + G_2\cdot \{Aq^3\}\{Aq^2\}\{A/q\}
\ee
Repeating the same reasoning for $N-4,5,\ldots,2r-1$ we iteratively deduce that
for arbitrary knot ${\cal K}$
\be
\boxed{
H_{[r]}^{\cal K} = \sum_{k=0}^r \frac{[r]!}{[k]![r-k]!} \cdot G_k^{\cal K}(A,q)\cdot
\left(\prod_{i=0}^{k-1}\{Aq^{r+i}\}\right)
\{A/q\}
}
\label{desymmK}
\ee
%\vspace{-0.4cm}
$$
= 1 + [r]\cdot G_1^{\cal K}(A,q)\cdot\{Aq^r\}\{A/q\} +
\frac{[r][r-1]}{[2]}\cdot G_2^{\cal K}(A,q)\cdot \{Aq^{r+1}\}\{Aq^r\}\{A/q\} + \ldots
$$
This is the generic form of symmetric differential expansion, suggested in \cite{Konodef}.
Transposed   version for antisymmetric representations is
\be
H_{[1^r]}^{\cal K} = \sum_{k=0}^r \frac{[r]!}{[k]![r-k]!} \cdot G_k^{\cal K}(A,q^{-1})\cdot\{Aq\}
\cdot \left(\prod_{i=0}^{k-1}\{A/q^{r+i}\}\right)
\ee

Original expansion (\ref{desymm41}) for the figure eight knot $4_1$  looks far more
restrictive.
Actually there are two levels of peculiarity: the coefficients $G_k$ are further factorized to
\be
G_k^{{\cal K}^{(d)}}\!\!(A,q) = F_k^{{\cal K}^{(d)}}\!\!(A,q)\cdot \prod_{i=1}^{k-d-1} \{Aq^{i-1}\}
\ee
with $d^{4_1}=0$, and the new coefficients $F_k^{4_1}(A,q)=1$.
Parameter $d^{\cal K}$ was named {\it defect} of the differential expansion in \cite{Konodef}
and it was conjectured that it is equal to the degree of the fundamental Alexander polynomial minus one
(polynomial should be taken in topological framing, where it is symmetric under the change
$q\longrightarrow q^{-1}$ and its degree is the maximal power of $q^2$, e.g.
$Al^{4_1}_{[1]} = H^{4_1}_{[1]}(A=1,q) = 1-\{q\}^2 = -q^2+3-q^{-2}$ has degree one and defect zero).
For polynomials of defect zero the first coefficient $G_1$ does not depend on $q$ --
such are all the twist knots,
as well as a slightly more general two-parametric  two-bridge family
called double-braid in \cite{MMMevo},
which needs to be studied  for extracting rectangular Racah matrices $\bar S$.

In the case of defect-zero knots one can say that the differential expansion is actually not
just in the differentials $D_n=\{Aq^n\}$, but in {\it quadratic} differentials
\be
Z^{(i)}_{[r]}=\{Aq^{r+i}\}\{Aq^{i-1}\}
\ee
i.e.
\be
\boxed{
H^{{\cal K}^{(0)}}_{[r]}\!\!(A,q) =
\sum_{k=0}^r \frac{[r]!}{[k]![r-k]!} \cdot F_k^{{\cal K}^{(0)}}\!\!(A,q)\cdot
\prod_{i=0}^{k-1}Z^{(i)}_{[r]}
}
\label{symdef0}
\ee
and one of the conjectures in the present paper
is that this property -- dependence on differentials through quadratic $Z^{(p)}_R$ --
survives for defect-zero knots for all rectangular diagrams $R=[r^s]$.

\section{Group theory restrictions for rectangular diagrams}

As we already know from (\ref{preHrs1}),
for rectangular diagrams $R=[r^s]$ the first term of differential expansion
is especially simple:
\be
H^{\cal K}_{[r^s]} -1 \ \sim \ \{Aq^r\}\{A/q^s\}
\label{Hrs1}
\ee
Further, from (\ref{preHrsN}) with $N=r+1$ and $N=s+1$ we get:
\be
H_{[r^s]}-H_{[r]} \sim \{A/q^{s+1}\}
\ \ \ \ \ \ \ \ \
{\rm and}
\ \ \ \ \ \ \ \ \
H_{[s^r]} - H_{[s]}\sim \{A/q^{r+1}\} \ \Longleftrightarrow \
H_{[r^s]}-H_{[1^s]} \sim \{Aq^{r+1}\}
\label{Hrsrp1}
\ee
from which we deduce that
\be
H^{\cal K}_{[r^s]} = 1 + \{Aq^r\}\{A/q^s\}\Big([r][s]G_1^{\cal K} + O(D)\Big)
\label{HrsG1}
\ee
with the same $G_1^{\cal K}(A,q)$ as in (\ref{desymmK}).

A much simpler corollary of (\ref{Hrsrp1}) is that simply the $r+2$-th and all further
terms of the differential expansion are divisible by $\{A/q^{s+1}\}$ and
$s+2$-th and further -- by $\{Aq^{r+1}\}$.
This simply follows from the assumption??? that $H_{[r]}$ and $H_{[1^s]}$ contain
respectively $r+1$ and $s+1$ different powers of the differentials.
To this one can add similar statements for higher $N$ -- and this already provides
somewhat powerful restrictions, which are further enhanced for defect-zero knots
by the conjecture of $Z_{[r^s]}$-dependence.

For $R=[22]$ we get in this way:
\be
\begin{array}{ccccccccc}
{\rm degree} & 0 && 1 &2 && 3 & 4  \\
\\
H_{[22]}^{\cal K}\ = & 1 &  + \{Aq^2\}\{A/q^2\}\cdot & \Big([2]^2 G_1 +  & ?\cdot \{\ ?\ \} &
+ \{Aq^3\}\{A/q^3\}\cdot &\Big( \ [2]^2\cdot ? & + ?\cdot \{\ ? \}  \Big)\Big)
\\ \\
& & \updownarrow \ (\ref{Hrs1}) &  \updownarrow \ (\ref{HrsG1}) && \updownarrow  \ (\ref{Hrsrp1}) &
\\ \\
H_{[22]}^{{\cal K}^{(0)}}\ = & 1 & + \ \ \ \ ^{(0)}\,\cdot & \Big([2]^2 F_1 +  & ?\cdot Z_{[22]}^{(?)} &
+ Z_{[22]}^{(1)}Z_{[22]}^{(-1)}\cdot & \Big( \ [2]^2\cdot ? & + ?\cdot Z_{[22]}^{(?)}\Big)\Big)
\end{array}
\label{22pred}
\ee
We see that in general group theory restrictions leave undetermined just two differential structures
and three coefficients, while in the case of defect zero the differential structures are almost fixed.
Indeed,  the transposition symmetry of the diagram $[22]$ requires the sets of superscripts $?$
in the two undetermined terms to be symmetric.
Since the last term has combinatorial multiplicity (binomial coefficient) one,
the only choice is $?=0$.
In the middle term the most natural choice would be $?=\pm 1$, so that

{\footnotesize
\be
H_{[22]}^{{\cal K}^{(0)}} \ \stackrel{?}{=} \ 1 +  \ Z_{[22]}^{(0)}\cdot  \left([2]^2 F_1(A,q) +
 [3] \Big(\tilde F_2(A,q)\cdot Z_{[22]}^{(1)} + \tilde F_2(A,q^{-1})\cdot Z_{[22]}^{(-1)}\Big)
 \ + Z_{[22]}^{(1)}Z_{[22]}^{(-1)}\cdot  \Big( \ [2]^2\tilde F_3(A,q)  +
 \tilde F_4(A,q) \cdot Z_{[22]}^{(0)}\Big)\right)
\nn
\ee
}
It remains to determine $\tilde F_2$ and $q\longleftrightarrow q^{-1}$ symmetric $\tilde F_3$ and $\tilde F_4$.
%?????It is, perhaps, not a very big surprise that actually
%$\tilde F_2=F_2$, $\tilde F_3=F_3(A,q=1)$ and $\tilde F_4=F_4(A,q=1)$???. ???
One can hope that they are made from $F_{2,3,4}$, describing the first four symmetric representations
(actually, in the case of $[22]$ the substitutions $q\longrightarrow q^0, q^{-1}$ can be sufficient).

For $R=[33]=[3^2]$ we have relations
\be
H_{[33]} - \underbrace{1}_{{\rm degree} \ 0} \sim \{A/q^2\}
& \Longleftarrow & [33] \ \stackrel{SU(2)}{\cong} \ \emptyset
\label{33deg0a}\\
H_{[33]} - \underbrace{1}_{{\rm degree} \ 0} \sim \{Aq^3\}
& \Longleftarrow & [222]=\widetilde{[33]} \ \stackrel{SU(3)}{\cong} \ \emptyset
\label{33deg0b} \\
H_{[33]} - \underbrace{H_{[11]}}_{{\rm degree} \ 2} \sim \{Aq^4\}
& \Longleftarrow & [222]=\widetilde{[33]} \ \stackrel{SU(4)}{\cong} \
[2] = \widetilde{[11]}
\label{33deg2}\\
H_{[33]} - \underbrace{H_{[3]}}_{{\rm degree} \ 3} \sim \{A/q^3\}
& \Longleftarrow & [33] \ \stackrel{SU(3)}{\cong} \ [3]
\label{33deg3}\\
H_{[33]} - \underbrace{H_{[22]}}_{{\rm degree} \ 4} \sim \{Aq^5\}
& \Longleftarrow &
[222]=\widetilde{[33]} \ \stackrel{SU(3)}{\cong} \ [22]=\widetilde{[22]}
\label{33deg4}
\ee
what implies that

\bigskip

\centerline{
{\footnotesize
$
\begin{array}{cccccccccccc}
{\rm degree} & 0 && 1 &2 && 3 && 4&&5& 6 \\
\\
H_{[33]}^{\cal K}\ = & 1
&  + \{Aq^3\}\{A/q^2\}\cdot & \Big([3][2] G_1 +  & ?\cdot \{\ ?\ \}
&+ \{Aq^4\}\cdot &\Big( \  ? \cdot \{\ ?\} & + \{A/q^3\}\cdot & \Big( \{\ ?\  \}
&+ \{Aq^5\}&\cdot\Big(\{\ ? \ \}
&+ ? \cdot \{ \ ?\ \} \{\ ?\ \}   \Big)\Big)\Big)\Big)
\\ \\
& & \updownarrow \ (\ref{33deg0a}) \ \& \ (\ref{33deg0b}) &
%\updownarrow \ (\ref{HrsG1})
&& \updownarrow  \ (\ref{33deg2}) &
& \updownarrow \ (\ref{33deg3}) &
& \updownarrow \ (\ref{33deg4}) &
\\ \\
H_{[33]}^{{\cal K}^{(0)}}\ = & 1
& + \ \ \ \ Z_{[33]}^{(0)}\cdot & \Big([3][2] F_1 +
& ?\cdot Z_{[33]}^{(?)} &
+ Z_{[33]}^{(1)}\cdot & \Big( ? \cdot Z_{[33]}^{(?)} & +   Z_{[33]}^{(-1)}\cdot
&\Big( ? \cdot Z_{[33]}^{(?)}
&+ Z_{[33]}^{(2)}\cdot &\Big( ? \cdot Z_{[33]}^{(?)}
& +? \cdot Z_{33}^{(?)}Z_{33}^{(?)}  \Big)\Big)   \Big)\Big)
\end{array}
$
}}

\bigskip

For $R=[44]=[4^2]$ the  relations are:
\be
H_{[44]} - \underbrace{1}_{{\rm degree} \ 0} \sim \{A/q^2\}
& \Longleftarrow & [33] \ \stackrel{SU(2)}{\cong} \ \emptyset
\label{44deg0a}\\
H_{[44]} - \underbrace{1}_{{\rm degree} \ 0} \sim \{Aq^4\}
& \Longleftarrow & [2222]=\widetilde{[44]} \ \stackrel{SU(4)}{\cong} \ \emptyset
\label{44deg0b} \\
H_{[44]} - \underbrace{H_{[11]}}_{{\rm degree} \ 2} \sim \{Aq^5\}
& \Longleftarrow & [2222]=\widetilde{[44]} \ \stackrel{SU(5)}{\cong} \
[2] = \widetilde{[11]}
\label{44deg2}\\
H_{[44]} - \underbrace{H_{[22]}}_{{\rm degree} \ 4} \sim \{Aq^6\}
& \Longleftarrow & [2222]=\widetilde{[44]} \ \stackrel{SU(6)}{\cong} \
[22] = \widetilde{[22]}
\label{44deg4a}\\
H_{[44]} - \underbrace{H_{[4]}}_{{\rm degree} \ 4} \sim \{A/q^3\}
& \Longleftarrow & [44] \ \stackrel{SU(3)}{\cong} \ [4]
\label{44deg4b}\\
H_{[44]} - \underbrace{H_{[33]}}_{{\rm degree} \ 6} \sim \{Aq^7\}
& \Longleftarrow &
[2222]=\widetilde{[44]} \ \stackrel{SU(7)}{\cong} \ [222]=\widetilde{[333]}
\label{44deg6}
\ee
and they imply

\bigskip

\centerline{
{\tiny
$
\begin{array}{cccccrcccccccccc}
{\rm degree} & 0 && 1 &2 && 3 & 4&&5& 6&& 7&8 \\
\\
H_{[44]}^{\cal K} \ = & 1 \!\!\!\!\!\!\!
&  + \{Aq^4\}\{A/q^2\}\cdot\!\!\!\!\!\!\! & \Big([4][2] G_1 +  & ?\cdot \{\ ?\ \}
&+ \{Aq^5\}\cdot \!\!\!\!\!\!\!&\Big( \  ? \cdot \{\ ?\} & + ?\cdot \{ \ ? \ \}\{ \ ? \ \}
&+ \{Aq^6\}\{A/q^3\}\cdot\!\!\!\!\!\!\!\!\!\! & \Big( \{\ ?\  \} & + ? \cdot  \{ \ ? \ \}\{ \ ? \ \}
&+ \{Aq^7\}\cdot\!\!\!\!\!\!\!&\Big(?\cdot \{\ ? \ \}\{\ ? \ \}
&+ ? \cdot \{ \ ?\ \} \{\ ?\ \}   \Big)\Big)\Big)\Big)
\\ \\
& & \updownarrow \ (\ref{44deg0a}) \ \& \ (\ref{44deg0b})\!\!\!\!\!\!\!\!\!\!\!\!\!\! &
%\updownarrow \ (\ref{HrsG1})
&& \updownarrow  \ (\ref{44deg2})\!\!\!\!\!\!\! &
&& \updownarrow \ (\ref{44deg4a}) \ \& \ (\ref{44deg4b})\!\!\!\!\!\!\!&
  &
&& \updownarrow \ (\ref{44deg6}) \!\!\!\!\!\!\!&
\\ \\
H_{[44]}^{{\cal K}^{(0)}}\ = & 1  \!\!\!\!\!\!\!
& + \ \ \ \ Z_{[44]}^{(0)}\cdot \!\!\!\!\!\!\!\!\!\!\!\!\!\! & \Big([4][2] F_1 +
& ?\cdot Z_{[44]}^{(?)} &
+ Z_{[44]}^{(1)}\cdot\!\!\!\!\!\!\!\! & \Big( ? \cdot Z_{[44]}^{(?)} & + ? \cdot Z_{[44]}^{(?)} Z_{[44]}^{(?)}
&+  Z_{[44]}^{(+2)}Z_{[44]}^{(-1)}\cdot\!\!\!\!\!\!\!
&\Big( ? \cdot Z_{[44]}^{(?)} & + ? \cdot Z_{[44]}^{(?)} Z_{[44]}^{(?)}
&+ Z_{[44]}^{(3)}\cdot \!\!\!\!\!\!\!&\Big( ? \cdot Z_{[44]}^{(?)}Z_{44}^{(?)}
& +? \cdot Z_{44}^{(?)}Z_{44}^{(?)} Z_{44}^{(?)} \Big)\Big)   \Big)\Big)
\end{array}
$
}}

\bigskip

Clearly, the number of undetermined structure is increasing with increase of $r$.
Still, it turns enough to reveal the structure of formulas and guess the answer,
when combined with information from another source, as we discuss in the next subsection.???

\section{Trefoil in rectangular representations}

Trefoil $3_1$ is a torus knot, therefore its HOMFLY is known in arbitrary representation
from the Rosso-Jones formula \cite{RJ}
and colored {\it hyper}polynomials -- from its straightforward generalization
\cite{AgSha,DMMSS,Che}.
Since rectangular representations do not suffer from the multiplicity problem,
superpolynomials for them presumably coincide with hyperpolynomials.
Moreover, there is a straightforward 4-graded generalization \cite{GGS,arthdiff}.

What is important for our purposes, trefoil is the only torus knot with defect zero,
thus it provides unvaluable information for generalizations
of the simplest type (\ref{symdef0}) of differential expansion to non-(anti)symmetric representations $R$.
In this paper we use it to find the knot-dependent coefficients in (\ref{22pred})
and its more complicated analogues  for ${\cal K}=3_1$.
After that we conjecture how they are modified for $4_1$ (this is easy).
In the future one can attempt generalizations to
other twist and, finally, double-braid knots -- what can be a far less reliable speculation.
Still the risk would pay for it -- from these conjectures one will deduce exclusive Racah matrices,
calculate colored HOMFLY for {\it arbitrary} arborescent knots,
and make new checks, involving arborescent torus knots: two-strand, $8_{19}$ and $???$.

\subsection{Representation $R=[22]=[2^2]$}

This is the case, where all arborescent knots were already exhaustively analyzed in \cite{mmms3},
based on a rigorous calculation of \cite{mmms2} for {\it inclusive} Racah matrices.
We now reproduce (some of) these results by the differential expansion method.

\be
\begin{array}{ccccccccc}
{\rm degree} & 0 && 1 &2 && 3 & 4  \\
\\
H_{[22]}^{\cal K}\ = & 1 &  + \{Aq^2\}\{A/q^2\}\cdot & \Big([2]^2 G_1 +  & ?\cdot \{\ ?\ \} &
+ \{Aq^3\}\{A/q^3\}\cdot &\Big( \ [2]^2\cdot ? & + ?\cdot \{\ ? \}  \Big)\Big)
\\ \\
& & \updownarrow \ (\ref{Hrs1}) &  \updownarrow \ (\ref{HrsG1}) && \updownarrow  \ (\ref{Hrsrp1}) &
\\ \\
H_{[22]}^{{\cal K}^{(0)}} \ = & 1 & + \ \ \ \ Z_{[22]}^{(0)}\,\cdot & \Big([2]^2 F_1 +  & ?\cdot Z_{[22]}^{(?)} &
+ Z_{[22]}^{(1)}Z_{[22]}^{(-1)}\cdot & \Big( \ [2]^2\cdot ? & + ?\cdot Z_{[22]}^{(?)}\Big)\Big)
\end{array}
\ee

\noindent
On the other hand, this should be an expansion of   the true Rosso-Jones answer,
and simple adjustment allows to substitute question signs by the full-fledged formula:
\be
H_{[22]}^{3_1} = 1  \underline{- [2]^2\, A^2 \,   Z_{[22]}^{(0)}}
+ [3]\,\Big( q^2A^4   Z_{[22]}^{(+1)} + q^{-2}A^4 Z_{[22]}^{(-1)}\Big) \underline{Z_{[22]}^{(0)}}
-\underline{[2]^2}\, A^6 \, \underline{Z_{[22]}^{(+1)}Z_{[22]}^{(0)}Z_{[22]}^{(-1)}}
+ A^8 \, \underline{Z_{[22]}^{(+1)}\big( Z_{[22]}^{(0)}\big)^2 Z_{[22]}^{(-1)}}
\nn
\ee
Underlined are the elements, {\it prescribed} by the group theory constraints
(\ref{22pred}).

Erasing all the coefficients $F_k^{3_1} = (-)^k q^{k(k-1)}A^{2k} \ \
\longrightarrow  \ \ F_k^{4_1}  = 1$, we obtain
\be
H_{[22]}^{4_1} = 1 + [2]^2\,    Z_{[22]}^{(0)}
+ [3]\,\Big(   Z_{[22]}^{(+1)} +   Z_{[22]}^{(-1)}\Big) Z_{[22]}^{(0)}
-[2]^2\,   Z_{[22]}^{(+1)}Z_{[22]}^{(0)}Z_{[22]}^{(-1)}
 + Z_{[22]}^{(+1)}\big(Z_{[22]}^{(0)}\big)^2Z_{[22]}^{(-1)}
\ee
what is the right answer, {\it derived} in \cite{mmms2}.

%Presumably for arbitrary defect-zero knot we should get
%
%{\footnotesize
%\be
%H_{[22]}^{{\cal K}^{(0)}} \ \stackrel{?}{=} \ 1 +  \ Z_{[22]}^{(0)}\cdot  \left([2]^2 F_1(A,q) +
% [3] \Big(\tilde F_2(A,q)\cdot Z_{[22]}^{(1)} + \tilde F_2(A,q^{-1})\cdot Z_{[22]}^{(-1)}\Big)
% \ + Z_{[22]}^{(1)}Z_{[22]}^{(-1)}\cdot  \Big( \ [2]^2\tilde F_3(A,q)  +
% \tilde F_4(A,q) \cdot Z_{[22]}^{(0)}\Big)\right)
%\nn
%\ee
%}
%Moreover, one can hope that  $\tilde F_{2,3,4}(A,q)$  are  made from
%$F_{2,3,4}(A,q)$, describing differential expansion in the first four symmetric representations.

\subsection{Representation $R=[33]=[3^2]$}

This time the group-theory-prescribed structure is

\bigskip

\centerline{
{\footnotesize
$
\begin{array}{cccccccccccc}
{\rm degree} & 0 && 1 &2 && 3 && 4&&5& 6 \\
\\
H_{[33]}^{\cal K}\ = & 1
&  + \{Aq^3\}\{A/q^2\}\cdot & \Big([3][2] G_1 +  & ?\cdot \{\ ?\ \}
&+ \{Aq^4\}\cdot &\Big( \  ? \cdot \{\ ?\} & + \{A/q^3\}\cdot & \Big( \{\ ?\  \}
&+ \{Aq^5\}&\cdot\Big(\ ?\cdot \{\ ? \ \}
&+ ? \cdot \{ \ ?\ \} \{\ ?\ \}   \Big)\Big)\Big)\Big)
\\ \\
& & \updownarrow \ (\ref{33deg0a}) \ \& \ (\ref{33deg0b}) &
%\updownarrow \ (\ref{HrsG1})
&& \updownarrow  \ (\ref{33deg2}) &
& \updownarrow \ (\ref{33deg3}) &
& \updownarrow \ (\ref{33deg4}) &
\\ \\
H_{[33]}^{{\cal K}^{(0)}}\ = & 1
& + \ \ \ \ Z_{[33]}^{(0)}\cdot & \Big([3][2] F_1 +
& ?\cdot Z_{[33]}^{(?)} &
+ Z_{[33]}^{(1)}\cdot & \Big( ? \cdot Z_{[33]}^{(?)} & +   Z_{[33]}^{(-1)}\cdot
&\Big( ? \cdot Z_{[33]}^{(?)}
&+ Z_{[33]}^{(2)}\cdot &\Big( ? \cdot Z_{[33]}^{(?)}
& +? \cdot Z_{33}^{(?)}Z_{33}^{(?)}  \Big)\Big)   \Big)\Big)
\end{array}
$
}}

\bigskip

\noindent
Again, it is not too difficult to convert the Rosso-Jones answer to this form:
\be
H_{[33]}^{3_1} = 1 \underline{ - [3][2]\, A^2 \,  Z_{[33]}^{(0)}}
+  \Big([3]^2\,q^2A^4\,   Z_{[33]}^{(+1)} + \frac{[3][4]}{[2]} q^{-2}A^4\, Z_{[33]}^{(-1)}\Big)
\underline{Z_{[33]}^{(0)}}
-  \Big([4]\, q^6A^6 \, Z_{[33]}^{(+2)} + [4][2]^2\,A^6\,Z_{[33]}^{(-1)}   \Big)
\underline{Z_{[33]}^{(+1)}Z_{[33]}^{(0)}}
+ \nn \\
\!\!\!\!\!\!\!\!\!\!\!
+\Big([3]^2\,q^4A^8 \, Z_{[33]}^{(+2)}+ \frac{[3][4]}{[2]}\,A^8\,Z_{[33]}^{(0)}\Big)
\underline{Z_{[33]}^{(+1)}Z_{[33]}^{(0)}Z_{[33]}^{(-1)}}
- [3][2]\,q^4A^{10}\, \underline{Z_{[33]}^{(+2)} Z_{[33]}^{(+1)}}
\big(\underline{Z_{[33]}^{(0)}}\big)^2 \underline{Z_{[33]}^{(-1)}}
+ q^6A^{12}\,
\underline{Z_{[33]}^{(+2)}} \big(\underline{Z_{[33]}^{(+1)} Z_{[33]}^{(0)}}\big)^2
\underline{Z_{[33]}^{(-1)}}
\nn
\ee
Note that prescribed (underlined) in the last two terms are all the factors, but {\it not}
their squares.
Again, erasing the coefficients, we conjecture that
\be
H_{[33]}^{4_1}\ \stackrel{?}{=}\ 1 + [3][2]\,    Z_{[33]}^{(0)}
+  \Big([3]^2   Z_{[33]}^{(+1)} + \frac{[3][4]}{[2]}  Z_{[33]}^{(-1)}\Big) Z_{[33]}^{(0)}
+  \Big([4]  \, Z_{[33]}^{(+2)} + [4][2]^2 \,Z_{[33]}^{(-1)}   \Big) Z_{[33]}^{(+1)}Z_{[33]}^{(0)}
+ \nn \\
+\Big([3]^2 \, Z_{[33]}^{(+2)}+ \frac{[3][4]}{[2]} \,Z_{[33]}^{(0)}\Big)
Z_{[33]}^{(+1)}Z_{[33]}^{(0)}Z_{[33]}^{(-1)}
+ [3][2]  Z_{[33]}^{(+2)} Z_{[33]}^{(+1)}\big(Z_{[33]}^{(0)}\big)^2 Z_{[33]}^{(-1)}
+Z_{[33]}^{(+2)} \big(Z_{[33]}^{(+1)} Z_{[33]}^{(0)}\big)^2 Z_{[33]}^{(-1)}
\ee

\subsection{Representation $R=[44] = [4^2]$}

A somewhat tedious analysis of the trefoil HOMFLY in this case brings it to the form:

{\footnotesize
\be
H_{[44]}^{3_1} = 1 \underline{ - [4][2]\, A^2 \,  Z_{[44]}^{(0)}}
+  \Big(\frac{[3]^2[4]}{[2]}\,q^2A^4\,   Z_{[44]}^{(+1)} + \frac{[4][5]}{[2]} q^{-2}A^4\, Z_{[44]}^{(-1)}\Big)
\underline{Z_{[44]}^{(0)}}
-  \Big([4]^2\, q^6A^6 \, Z_{[44]}^{(+2)} + {[5][4][2]}\,A^6\,Z_{[44]}^{(-1)}   \Big)
\underline{Z_{[44]}^{(+1)}Z_{[44]}^{(0)}}
+ \nn \\
+\Big([5]\,q^{12} A^8 \, Z_{[44]}^{(+3)}Z_{[44]}^{(+2)}+ [5][3]^2\,q^4A^8 \, Z_{[44]}^{(+2)}Z_{[44]}^{(-1)}
+ \frac{[5][4]^2}{[2]^2}\,A^8\,Z_{[44]}^{(0)}Z_{[44]}^{(-1)}\Big)
\underline{Z_{[44]}^{(+1)}Z_{[44]}^{(0)}}
- \nn \\
-\Big([4]^2\,q^{10}A^{10}\,Z_{[44]}^{(+3)}+[5][4][2]\,q^4A^{10}\, Z_{[44]}^{(0)}\Big)
\underline{Z_{[44]}^{(+2)} Z_{[44]}^{(+1)}Z_{[44]}^{(0)} Z_{[44]}^{(-1)}}
+ \Big(\frac{[3]^2[4]}{[2]}\,q^{10}A^{12}\,Z_{[44]}^{(+3)} + \frac{[5][4]}{[2]}\,q^6A^{12}\,Z_{[44]}^{(+1)}\Big)
\underline{Z_{[44]}^{(+2)} Z_{[44]}^{(+1)}}\big(\underline{Z_{[44]}^{(0)}}\big)^2 \underline{Z_{[44]}^{(-1)}}
- \nn \\
- [4][2]\,q^{12}A^{14}
\underline{Z_{[44]}^{(+3)}Z_{[44]}^{(+2)}}\big(\underline{ Z_{[44]}^{(+1)}Z_{[44]}^{(0)}}\big)^2
\underline{Z_{[44]}^{(-1)}}
+ q^{16}A^{16}\,
Z_{[44]}^{(+3)}\big(\underline{Z_{[44]}^{(+2)} Z_{[44]}^{(+1)} Z_{[44]}^{(0)}}\big)^2
\underline{Z_{[44]}^{(-1)}}
\ee
}
Then the conjecture for the figure eight knot is

{\footnotesize
\be
H_{[44]}^{4_1}\ \stackrel{?}{=}\ 1  + [4][2]\,    Z_{[44]}^{(0)}
+  \Big(\frac{[3]^2[4]}{[2]} \,   Z_{[44]}^{(+1)} + \frac{[4][5]}{[2]}  \, Z_{[44]}^{(-1)}\Big)
 Z_{[44]}^{(0)}
+  \Big([4]^2\,   Z_{[44]}^{(+2)} + {[5][4][2]}\, Z_{[44]}^{(-1)}   \Big)
 {Z_{[44]}^{(+1)}Z_{[44]}^{(0)}}
+ \nn \\
+\Big([5]  \, Z_{[44]}^{(+3)}Z_{[44]}^{(+2)}+ [5][3]^2  \, Z_{[44]}^{(+2)}Z_{[44]}^{(-1)}
+ \frac{[5][4]^2}{[2]^2} \,Z_{[44]}^{(0)}Z_{[44]}^{(-1)}\Big)
 {Z_{[44]}^{(+1)}Z_{[44]}^{(0)}}
+ \nn \\
+\Big([4]^2\, Z_{[44]}^{(+3)}+[5][4][2] \, Z_{[44]}^{(0)}\Big)
 {Z_{[44]}^{(+2)} Z_{[44]}^{(+1)}Z_{[44]}^{(0)} Z_{[44]}^{(-1)}}
+ \Big(\frac{[3]^2[4]}{[2]} \,Z_{[44]}^{(+3)} + \frac{[5][4]}{[2]} \,Z_{[44]}^{(+1)}\Big)
 {Z_{[44]}^{(+2)} Z_{[44]}^{(+1)}}\big( {Z_{[44]}^{(0)}}\big)^2  {Z_{[44]}^{(-1)}}
+ \nn \\
+ [4][2]
 Z_{[44]}^{(+3)}Z_{[44]}^{(+2)}{\big( Z_{[44]}^{(+1)}} {Z_{[44]}^{(0)}}\big)^2  {Z_{[44]}^{(-1)}}
+
Z_{[44]}^{(+3)}\big( {Z_{[44]}^{(+2)}}  {Z_{[44]}^{(+1)} Z_{[44]}^{(0)}}\big)^2
 {Z_{[44]}^{(-1)}}
\ee
}
Now the structure of the answer is getting clear and we can attempt a more general conjecture.

\subsection{ Representation $R=[rr]=[r^2]$}

For this we need to guess general formulas for the coefficients.
This actually requires additional insights from Rosso-Jones answers for higher $r$ --
non of them can actually be handled by itself, but alltogether they provide
sufficient information for an educated guesswork.
Once the result emerges, it looks obviously true:

\bigskip

{\footnotesize
\be
H_{[rr]}^{3_1} = 1 \ -\  A^2\, [2][r]\,     Z^{(0)} \
+ \  A^4\Big(q^2 \,\frac{[3][r][r-1]}{[2]} \, Z^{(1)} +  q^{-2} \,\frac{[r][r+1]}{[2]}\, Z^{(-1)}\Big)Z^{(0)}
- \nn \\
-A^6 \Big(q^6\,\frac{[4][r]!}{[3]![r-3]!}\, Z^{(2)}
+ q^0\, \frac{[2][r-1][r][r+1]}{[3]}  \,Z^{(-1)}\Big) Z^{(1)}Z^{(0)}
+ \nn \\
+ A^8 \Big(  q^{12} \,\frac{[5][r]!}{[4]![r-4]!} \, Z^{(3)}Z^{(2)}
+ q^4\, \frac{[3][r+1]!}{[2][4][r-3]!} \,Z^{(2)}Z^{(-1)} +
q^0\,\frac{[r-1][r]^2[r+1]}{[2]^2[3]}\,Z^{(0)}Z^{(-1)}\Big) Z^{(1)}Z^{(0)}
- \nn \\
-A^{10}\Big(q^{20}\,\frac{[6][r]!}{[5]![r-5]!}\, Z^{(4)}Z^{(3)}
+ q^{10}\,\frac{[4][r+1]!}{[2][3][5][r-4]!}\,Z^{(3)}Z^{(-1)}
+ q^4\,\frac{[r][r+1]!}{[3][4][r-3]!}\,Z^{(0)}Z^{(-1)}\Big)
Z^{(2)}Z^{(1)}Z^{(0)}
+ \nn
\ee
\be
\!\!\!\!\!\!\!\!\!\!\!\!\!\!\!\!\!\!\!\!\!\!\!\!\!\!\!\!\!\!\!\!\!\!\!
+A^{12}\Big(q^{30}\,\frac{[7][r]!}{[6]![r-6]!}\,Z^{(5)}Z^{(4)}Z^{(3)}
+ q^{18}  \frac{[5][r+1]!}{[4]! [6][r-5]!}\,Z^{(4)}Z^{(3)}Z^{(-1)}
+ q^{10}\,\frac{[3][r][r+1]!}{[2]^2[4][5][r-4]!}\,Z^{(3)}Z^{(0)}Z^{(-1)}
+ q^6\, \frac{[r-1][r][r+1]!}{[2][3][4]![r-3]!}\,Z^{(1)}Z^{(0)}Z^{(-1)} \Big)Z^{(2)}Z^{(1)}Z^{(0)}
-\nn
\ee
\be
\!\!\!\!\!\!\!\!\!\!\!\!\!\!\!\!\!\!\!\!\!\!\!\!\!\!\!\!\!\!\!\!\!\!\!\!\!\!\!\!\!\!\!
-A^{14}\Big(q^{42}\,\frac{[8][r]!}{[7]![r-7]!}\,Z^{(6)}Z^{(5)}Z^{(4)}
+ q^{28}\,\frac{[6][r+1]!}{[5]![7][r-6]!}Z^{(5)}Z^{(4)}Z^{(-1)}
+ q^{18}\, \frac{[4]^2[r][r+1]!}{[2][6]![r-5]!}\,Z^{(4)}Z^{(0)} Z^{(-1)}
+ q^{12}\,\frac{[2] [r-1][r][r+1]!}{ [3][5]![r-4]!}\,Z^{(1)}Z^{(0)}Z^{(-1)}\Big) Z^{(3)}Z^{(2)}Z^{(1)}Z^{(0)}
+ \nn
\ee
\be
\!\!\!\!\!\!\!\!\!\!\!\!\!\!\!\!\!\!\!\!\!\!\!\!\!\!\!\!\!\!\!\!\!\!\!\!\!\!\!\!\!\!
+ A^{16}\Big(q^{56}\,\frac{[9][r]!}{[8]![r-8]!}\,Z^{(7)}Z^{(6)}Z^{(5)}Z^{(4)}
+ q^{40}\frac{[7][r+1]!}{[6]![8][r-7]!}Z^{(6)}Z^{(5)}Z^{(4)}Z^{(-1)}
+ q^{28} \frac{[5]^2[r][r+1]!}{[2][7]![r-6]!} Z^{(5)}Z^{(4)}Z^{(0)}Z^{(-1)}
+\nn \\
+ q^{20} \frac{[3]^2[r-1][r][r]!}{[2][3][6]![r-5]!} Z^{(4)}Z^{(1)}Z^{(0)}Z^{(-1)}
+ q^{16}\,\frac{[r-2][r-1][r][r+1]!}{[2][3][4][4]![r-4]!}\,Z^{(2)}Z^{(1)}Z^{(0)}Z^{(-1)}
\Big)Z^{(3)}Z^{(2)}Z^{(1)}Z^{(0)}
 -
 \nn
 \ee
 \be
 \!\!\!\!\!\!\!\!\!\!\!\!\!\!\!\!\!\!\!\!\!\!\!\!\!\!\!\!\!\!\!\!\!
-A^{18}\Big(q^{72}\,\frac{[10][r]!}{[9]![r-9]!}\,Z^{(8)}Z^{(7)}Z^{(6)}Z^{(5)}
 +q^{54} \frac{[8][r]!}{[7]![9][r-8]!}Z^{(7)}Z^{(6)}Z^{(5)}Z^{(-1)}
 + q^{40}\frac{[6]^2[r][r+1]!}{[2][8]![r-7]!}Z^{(6)}Z^{(5)}Z^{(0)}Z^{(-1)}
 + \nn \\
 + q^{30}\frac{[4]^2[r-1][r][r]!}{[2][3][7]![r-6]!}Z^{(5)}Z^{(1)}Z^{(0)}Z^{(-1)}
+ q^{24}\frac{[2][r-2][r-1][r][r+1]!}{[3][4][5]![r-5]!}\,Z^{(2)}Z^{(1)}Z^{(0)}Z^{(-1)}
\Big) Z^{(4)}Z^{(3)}Z^{(2)}Z^{(1)}Z^{(0)}
+ \nn
\ee
\be
%\!\!\!\!\!\!\!\!\!\!\!\!\!\!\!\!\!\!\!\!\!\!\!\!\!\!\!\!\!\!\!\!\!
%\!\!\!\!\!\!\!\!\!\!\!\!\!\!\!\!\!\!\!\!\!\!\!\!\!\!\!\!\!\!\!\!\!
+ A^{20}\Big( q^{90}\,\frac{[11][r]!}{[10]![r-10]!}\, Z^{(9)} Z^{(8)}Z^{(7)}Z^{(6)}Z^{(5)}
+q^{70}\frac{[9][r+1]!}{[8]![10][r-9]!}  Z^{(8)}Z^{(7)}Z^{(6)}Z^{(5)}Z^{(-1)}
+ q^{54} \frac{[7]^2[r][r+1]!}{[2][9]![r-8]!} Z^{(7)}Z^{(6)}Z^{(5)}Z^{(0)}Z^{(-1)}
+\nn \\
%\!\!\!\!\!\!\!\!\!\!\!\!\!\!\!\!\!\!\!\!\!\!\!\!\!\!\!\!\!\!\!\!\!
+ q^{42}\frac{[5]^2[r-1][r][r]!}{[2][3][8]![r-7]!} Z^{(6)}Z^{(5)}Z^{(1)}Z^{(0)}Z^{(-1)}
+ q^{34} \frac{[3][r-2][r-1][r][r+1]!}{[2][4][6]![r-6]!} Z^{(5)}Z^{(2)}Z^{(1)}Z^{(0)}Z^{(-1)}
+\nn \\
+ q^{30}\frac{ [r]![r+1]!}{[5]![6]![r-4]![r-5]!}\,Z^{(3)}Z^{(2)}Z^{(1)}Z^{(0)}Z^{(-1)}
\Big) Z^{(4)}Z^{(3)}Z^{(2)}Z^{(1)}Z^{(0)} - \nn \\
-\ldots \ =
\nn
\ee
}
\be
\!\!\!\!\!\!\!\!
= \sum_{p=0}^{2r} (-)^p q^{p(p-1)} A^{2p} \left(\prod_{i=0}^{p-1} Z_{[rr]}^{(i )}\right)\cdot
\left(
\frac{[p+1][r]!}{[p]![r-p]!} + q^{-2p} \frac{[p-1]^2[r+1]!}{  [p]! [r-p+1]!}\frac{Z^{(-1)}}{Z^{(p-1)}}
+ q^{-4(p-1)} \frac{[p-3]^2[r][r+1]!}{[2][p-1]![r-p+2]!}\frac{Z^{(0)}Z^{(-1)}}{Z^{(p-1)}Z^{(p-2)}}
+\right.\nn
\ee
\be
\!\!\!\!\!\!\!\!\!\!\!\!\!\!
\left.  + q^{-6(p-2)} \frac{[p-5]^2[r][r-1][r+1]!}{[2][3][p-2]![r-p+3]!}
\frac{Z^{(1)}Z^{(0)}Z^{(-1)}}{Z^{(p-1)}Z^{(p-2)}Z^{(p-3)}}
+ q^{-8(p-3)} \frac{[p-7]^2[r][r-1][r-2][r+1]!}{[2][3][4][p-3]![r-p+4]!}
\frac{Z^{(2)}Z^{(1)}Z^{(0)}Z^{(-1)}}{Z^{(p-1)}Z^{(p-2)}Z^{(p-3)}Z^{(p-4)}} + \ldots  \right)
\nn
\ee
From these expressions it is clear, that
contributing to $H_{[rr]}^{3_1}$ are the $Z^{(r)}$-independent terms in the following pyramid, i.e.
lying over the $r$-th sub-diagonal:

\bigskip

\centerline{
{\footnotesize
$
\begin{array}{rrcccccccccccc}
1\cdot &1 \\
-A^2\cdot & Z^{(0)} \\
A^4 \cdot & Z^{(1)}Z^{(0)}  &\Big(1 &\oplus \frac{Z^{(-1)}}{Z^{(1)}} \Big) \\
-A^6 \cdot & Z^{(2)}Z^{(1)}Z^{(0)} & \Big( 1 &\oplus \frac{Z^{(-1)}}{Z^{(2)}}\Big) \\
A^8 \cdot & Z^{(3)}Z^{(2)}Z^{(1)}Z^{(0)} & \Big( 1 &\oplus \frac{Z^{(-1)}}{Z^{(3)}}
  &\oplus \frac{Z^{(0)}Z^{(-1)}}{Z^{(3)}Z^{(2)}}\Big) \\
-A^{10}\cdot   & Z^{(4)}Z^{(3)}Z^{(2)}Z^{(1)}Z^{(0)} & \Big( 1 &\oplus \frac{Z^{(-1)}}{Z^{(4)}}
  &\oplus \frac{Z^{(0)}Z^{(-1)}}{Z^{(4)}Z^{(3)}}\Big) \\
  A^{12}\cdot   & Z^{(5)}Z^{(4)}Z^{(3)}Z^{(2)}Z^{(1)}Z^{(0)} & \Big( 1 &\oplus \frac{Z^{(-1)}}{Z^{(5)}}
  &\oplus \frac{Z^{(0)}Z^{(-1)}}{Z^{(5)}Z^{(4)}}
  &\oplus \frac{Z^{(1)}Z^{(0)}Z^{(-1)}}{Z^{(5)}Z^{(4)}Z^{(3)}}\Big) \\
 - A^{14}\cdot   & Z^{(6)}Z^{(5)}Z^{(4)}Z^{(3)}Z^{(2)}Z^{(1)}Z^{(0)}
 & \Big( 1 &\oplus \frac{Z^{(-1)}}{Z^{(6)}}
  &\oplus \frac{Z^{(0)}Z^{(-1)}}{Z^{(6)}Z^{(5)}}
  &\oplus \frac{Z^{(1)}Z^{(0)}Z^{(-1)}}{Z^{(6)}Z^{(5)}Z^{(4)}}\Big) \\
  A^{16}\cdot   & Z^{(7)}Z^{(6)}Z^{(5)}Z^{(4)}Z^{(3)}Z^{(2)}Z^{(1)}Z^{(0)}
  & \Big( 1 &\oplus \frac{Z^{(-1)}}{Z^{(7)}}
  &\oplus \frac{Z^{(0)}Z^{(-1)}}{Z^{(7)}Z^{(6)}}
  &\oplus \frac{Z^{(1)}Z^{(0)}Z^{(-1)}}{Z^{(7)}Z^{(6)}Z^{(5)}}
  &\oplus \frac{Z^{(2)}Z^{(1)}Z^{(0)}Z^{(-1)}}{Z^{(7)}Z^{(6)}Z^{(5)}Z^{(4)}}\Big) \\
 -A^{18}\cdot   & Z^{(8)}Z^{(7)}Z^{(6)}Z^{(5)}Z^{(4)}Z^{(3)}Z^{(2)}Z^{(1)}Z^{(0)}
 & \Big( 1 &\oplus \frac{Z^{(-1)}}{Z^{(8)}}
  &\oplus \frac{Z^{(0)}Z^{(-1)}}{Z^{(8)}Z^{(7)}}&\oplus \frac{Z^{(1)}Z^{(0)}Z^{(-1)}}{Z^{(8)}Z^{(7)}Z^{(6)}}
  &\oplus \frac{Z^{(2)}Z^{(1)}Z^{(0)}Z^{(-1)}}{Z^{(8)}Z^{(7)}Z^{(6)}Z^{(5)}}\Big) \\
  A^{20}\cdot   & Z^{(9)}Z^{(8)}Z^{(7)}Z^{(6)}Z^{(5)}Z^{(4)}Z^{(3)}Z^{(2)}Z^{(1)}Z^{(0)}
  & \Big( 1 &\oplus \frac{Z^{(-1)}}{Z^{(9)}}
  &\oplus \frac{Z^{(0)}Z^{(-1)}}{Z^{(9)}Z^{(8)}}&\oplus \frac{Z^{(1)}Z^{(0)}Z^{(-1)}}{Z^{(9)}Z^{(8)}Z^{(7)}}
  &\oplus \frac{Z^{(2)}Z^{(1)}Z^{(0)}Z^{(-1)}}{Z^{(9)}Z^{(8)}Z^{(7)}Z^{(6)}}
  &\oplus \frac{Z^{(3)}Z^{(2)}Z^{(1)}Z^{(0)}Z^{(-1)}}{Z^{(9)}Z^{(8)}Z^{(7)}Z^{(6)}Z^{(5)}}\Big) \\
  \ldots
\end{array}
$
}}

\bigskip

\noindent
Because of the two-step edges at the right-hand side the number of such terms is always finite.
Direct sum sign $\oplus$ stands for omitted factors, made from quantum numbers and powers of $q$.
They are explicit in exact formula:
%and finally for the trefoil
\be
\boxed{
H_{[rr]}^{3_1}
= \sum_{p=0}^{2r} (-)^p q^{p(p-1)} A^{2p} \left(\prod_{i=0}^{p-1} Z_{[rr]}^{(i )}\right)\cdot
\sum_{k=0}^{ p/2 } q^{-2k(p+1-k)}  \frac{[p+1-2k]^2}{[k]![p+1-k]!}
\frac{[r]![r+1]!}{[r-k+1]! [r-p+k]!}  \prod_{i=1}^k \frac{Z^{(i-2)}}{Z^{(p-i)}}
}
\ee
Once again, the answer for the trefoil is known from the Rosso-Jones formula --
the goal of above manipulations was to convert it to the differential-expansion form,
where transition to the figure eight case is straightforward.
From this formula we get (conjecturally):
\be
\boxed{
H_{[rr]}^{4_1}
\ \stackrel{?}{=}\
\sum_{p=0}^{2r}  \left(\prod_{i=0}^{p-1} Z_{[rr]}^{(i )}\right)\cdot
\sum_{k=0}^{ p/2 }    \frac{[p+1-2k]^2}{[k]![p+1-k]!}
\frac{[r]![r+1]!}{[r-k+1]! [r-p+k]!}  \prod_{i=1}^k \frac{Z^{(i-2)}}{Z^{(p-i)}}
}
\label{41rr}
\ee

\subsection{Representation $R=[333] = [3^3]$}

This time group-theory restrictions are not very serious:
\be
H_{[333]}^{{\cal K}^{(0)}} = 1 + Z_{[333]}^{(0)}\Big([3]^2F_1 + \ ?\cdot Z +\ ? \cdot Z\cdot Z
+ Z_{[333]}^{(1)}Z_{[333]}^{(-1)}\Big(\ ?\cdot Z +\ ?\cdot Z\cdot Z +\
?\cdot Z\cdot\cdot Z\cdot Z + \nn \\
+ Z_{[333]}^{(2)}Z_{[333]}^{(-2)}\Big(\ ?\cdot Z\cdot Z +\
?\cdot Z\cdot\cdot Z\cdot Z +\ ?\cdot Z\cdot\cdot Z\cdot Z\cdot Z\Big)\Big)\Big)
\ee
but we already have enough experience to succeed almost without them.
The result is:

%{\footnotesize
\be
H_{[333]}^{3_1} = 1 -A^2\, [3]^2\, Z_{[333]}^{(0)}
+  A^4\,\frac{[4][3]^2}{[2]}\Big(q^2Z^{(1)}_{[333]} + q^{-2}Z^{(-1)}_{[333]}\Big)\,Z^{(0)}_{[333]}
-\nn\\
- A^6 \left(\frac{[5][4]}{[2]}
\Big(q^6 Z^{(2)}_{[333]}Z^{(1)}_{[333]}+q^{-6}Z^{(-2)}_{[333]}Z^{(-1)}_{[333]}\Big)Z^{(0)}_{[333]}+
[4]^2[2]^2Z^{(1)}_{[333]}Z^{(0)}_{[333]}Z^{(-1)}_{[333]}\right)
+ \nn \\
+ A^8\left( [3]^2[5]\Big(q^4Z^{(2)}_{[333]}+q^{-4}Z^{(-2)}_{[333]}\Big)
+ \frac{[3]^2[4]^2}{[2]^2}  Z^{(0)}_{[333]}  \right)Z^{(1)}_{[333]}Z^{(0)}_{[333]}Z^{(-1)}_{[333]}
- \nn \\
-A^{10}\left( [3]^2[5]
\Big(q^{4}Z^{(2)}_{[333]}+q^{-4}Z^{(-2)}_{[333]} \Big)Z^{(0)}_{[333]}
+ \frac{[3]^2[4]^2}{[2]^2} Z^{(2)}_{[333]}Z^{(-2)}_{[333]}  \right)Z^{(1)}_{[333]}Z^{(0)}_{[333]}Z^{(-1)}_{[333]}
+ \nn \\
%\!\!\!\!\!\!\!\!\!\!\!\!\!\!\!\!\!\!\!\!\!\!\!\!\!\!\!\!\!\!
+A^{12}\left(\frac{[5][4]}{[2]}
\Big(q^{6} Z^{(2)}_{[333]}Z^{(1)}_{[333]}
+q^{-6}Z^{(-2)}_{[333]}Z^{(-1)}_{[333]} \Big) +
[4]^2[2]^2Z^{(2)}_{[333]}   Z^{(-2)}_{[333]}\right)
Z^{(1)}_{[333]}\Big(Z^{(0)}_{[333]}\Big)^2Z^{(-1)}_{[333]}
- \nn \\
-A^{14}\,\frac{[3]^2[4]}{[2]}\Big(q^{2} Z^{(1)}_{[333]}   +
q^{-1} Z^{(-1)}_{[333]}   \Big)\,
Z^{(2)}_{[333]}Z^{(1)}_{[333]}\Big(Z^{(0)}_{[333]}\Big)^2Z^{(-1)}_{[333]}Z^{(-2)}_{[333]}
+ \nn
\ee
\vspace{-0.6cm}
\be
+\Big(A^{16}[3]^2 -A^{18}Z^{(0)}_{[333]}\Big)
Z^{(2)}_{[333]}\Big(Z^{(1)}_{[333]}Z^{(0)}_{[333]}Z^{(-1)}_{[333]}\Big)^2Z^{(-2)}_{[333]}
\ee
%}

\subsection{Representation $R=[444] = [4^3]$}

Similarly,
\be
H_{[444]}^{{\cal K}^{(0)}} = 1 + Z_{[444]}^{(0)}\Big([4][3]^2F_1 + \ ?\cdot Z +\ ? \cdot Z\cdot Z
+ Z_{[444]}^{(1)}\Big(\ ?\cdot Z\cdot Z
+ Z_{[444]}^{(-1)}\Big(\ ?\cdot Z\cdot Z +\ ?\cdot Z\cdot\cdot Z\cdot Z + \nn \\
+ Z_{[444]}^{(2)}\Big(\ ?\cdot Z\cdot Z\cdot Z +\ ?\cdot Z\cdot Z\cdot Z\cdot Z
+ Z_{[444]}^{(-2)}\Big(\ ?\cdot Z\cdot Z\cdot Z\cdot Z + \nn \\
+ Z_{[444]}^{(3)}\Big(\ ?\cdot Z\cdot Z\cdot Z\cdot Z +\ ?\cdot Z\cdot Z\cdot Z\cdot Z\cdot Z
+\ ?\cdot Z\cdot Z\cdot Z\cdot Z\cdot Z\cdot Z\Big)\Big)\Big)\Big)\Big)\Big)
\label{grth444}
\ee

\noindent
Though now factorizations are even less restrictive, they are "split" and in result the
related constraints appear
more frequently, thus facilitating adjustment of the coefficients.
The outcome is

{\footnotesize
\be
H_{[444]}^{3_1} = 1 -A^2\, [4][3]\, Z_{[444]}^{(0)}
+  A^4 Z^{(0)}_{[444]}
\left(q^2\frac{[4]^2[3]^2}{[2]^2}\,Z^{(1)}_{[444]} + q^{-2}\frac{[5][4][3]}{[2]}\,Z^{(-1)}_{[444]}\right)
-\nn\\
- A^6 Z^{(0)}_{[444]} \left(q^6\frac{[5][4]^2}{[2]}\, Z^{(2)}_{[444]}Z^{(1)}_{[444]}
+ [5][4]^2[2] \,Z^{(1)}_{[444]} Z^{(-1)}_{[444]}
+q^{-6}\frac{[6][5][4]}{[3][2]}\,Z^{(-2)}_{[444]}Z^{(-1)}_{[444]}\right)
 + \nn \\
+ A^8Z^{(1)}_{[444]}Z^{(0)}_{[444]}\left(q^{12}\frac{[6][5]}{[2]}Z^{(3)}_{[444]}Z^{(2)}_{[444]}
+q^4 [5]^2[3]^2Z^{(2)}_{[444]} Z^{(-1)}_{[444]}
+ \frac{[5][4]^3[3]}{[2]^3}  Z^{(0)}_{[444]} Z^{(-1)}_{[444]}
%+\right. \nn \\ \left.
+ q^{-4}\frac{[6][5][3]^2}{[2]}  Z^{(-1)}_{[444]}Z^{(-2)}_{[444]}\right)
-\nn \\
- A^{10} Z^{(1)}_{[444]}Z^{(0)}_{[444]}Z^{(-1)}_{[444]}\left(
q^{10}[4]^2[6]Z^{(3)}_{[444]}Z^{(2)}_{[444]}  +
q^4[5]^2[4][3]Z^{(2)}_{[444]}Z^{(0)}_{[444]} +
\frac{[6][4]^2[3]^2}{[2]^2} Z^{(2)}_{[444]}Z^{(-2)}_{[444]} +
q^{-4} \frac{[6][5][4][3]}{[2]} Z^{(-2)}_{[444]}Z^{(0)}_{[444]}
\right)
+ \nn \\ \nn \\
\!\!\!\!\!\!\!\!\!\!\!\!\!\!\!\!\!
+A^{12} Z^{(1)}_{[444]}Z^{(0)}_{[444]}Z^{(-1)}_{[444]}\left(
q^{10}\frac{[6][4][3]^3}{[2]^2} Z^{(3)}_{[444]}Z^{(2)}_{[444]}Z^{(0)}_{[444]}
+ q^6\frac{[5]^2[4]^2}{[2]^2} Z^{(3)}_{[444]}Z^{(2)}_{[444]}Z^{(-2)}_{[444]}
+ \right. \ \ \ \ \ \ \ \ \ \ \ \ \ \ \ \ \ \ \ \ \ \ \ \ \ \ \ \ \ \ \ \ \ \ \ \
  \nn \\ + \left.
q^6\frac{[5]^2[4]^2}{[2]^2} Z^{(2)}_{[444]}Z^{(1)}_{[444]}Z^{(0)}_{[444]}
+ \frac{[6][4]^3[2]^2}{[3]} Z^{(2)}_{[444]}Z^{(0)}_{[444]}Z^{(-2)}_{[444]}
+ q^{-6} \frac{[6][5]^2[4]}{[3][2]^2} Z^{(0)}_{[444]}Z^{(-1)}_{[444]}Z^{(-2)}_{[444]} \right)
- \nn \\ \nn \\
- A^{14} Z^{(2)}_{[444]}Z^{(1)}_{[444]}\Big(Z^{(0)}_{[444]}\Big)^2Z^{(-1)}_{[444]}\left(
q^{12}[6][4]^2 Z^{(3)}_{[444]}Z^{(1)}_{[444]}
+ q^6[5]^2[4][3] Z^{(3)}_{[444]}Z^{(-2)}_{[444]}
+  q^2\frac{[4]^2[3]^2}{[2]^2} Z^{(1)}_{[444]}Z^{(-2)}_{[444]}
+ q^{-2}\frac{[6][5][4][3]}{[2]} Z^{(-1)}_{[444]}Z^{(-2)}_{[444]}\right)
+ \nn \\
\!\!\!\!\!\!\!\!\!\!\! \!\!\!\!\!\!\!\!\!\!\! \!\!\!\!\!\!
+ A^{16} Z^{(2)}_{[444]}Z^{(1)}_{[444]}\Big(Z^{(0)}_{[444]}\Big)^2Z^{(-1)}_{[444]}\left(
q^{16}\frac{[6][5]}{[2]}Z^{(3)}_{[444]}Z^{(2)}_{[444]}Z^{(1)}_{[444]}
+ q^8[5]^2[3]^2 Z^{(3)}_{[444]}Z^{(1)}_{[444]}Z^{(-2)}_{[444]}
+ q^4\frac{[5][4]^3[3]}{[2]^3} Z^{(3)}_{[444]}Z^{(-1)}_{[444]}Z^{(-2)}_{[444]}
+ \frac{[6][5][3]^2}{[2]} Z^{(1)}_{[444]}Z^{(-1)}_{[444]}Z^{(-2)}_{[444]}
\right)
-\nn \\
- A^{18} Z^{(2)}_{[444]}\Big(Z^{(1)}_{[444]}Z^{(0)}_{[444]}\Big)^2Z^{(-1)}_{[444]}\left(
q^{12}\frac{[5][4]^2}{[2]} Z^{(3)}_{[444]}Z^{(2)}_{[444]}Z^{(-2)}_{[444]}
+ q^6[5][4]^2[2] Z^{(3)}_{[444]}Z^{(-1)}_{[444]}Z^{(-2)}_{[444]}
+ \frac{[6][5][4]}{[3][2]} Z^{(0)}_{[444]}Z^{(-1)}_{[444]}Z^{(-2)}_{[444]}\right)
+ \nn \\
+ A^{20} Z^{(3)}_{[444]}Z^{(2)}_{[444]}\Big(Z^{(1)}_{[444]}Z^{(0)}_{[444]}Z^{(-1)}_{[444]}\Big)^2
Z^{(-2)}_{[444]}\left(
q^{10}\frac{[4]^2[3]^2}{[2]^2} Z^{(2)}_{[444]} + q^6\frac{[5][4][3]}{[2]} Z^{(0)}_{[444]}\right)
- \nn \\
-A^{22}q^{10}[4][3]Z^{(3)}_{[444]}\Big(Z^{(2)}_{[444]}Z^{(1)}_{[444]}\Big)^2\Big(Z^{(0)}_{[444]}\Big)^3
\Big(Z^{(-1)}_{[444]}\Big)^2 Z^{(-2)}_{[444]}
 + A^{24}q^{12} Z^{(3)}_{[444]}\Big(Z^{(2)}_{[444]}\Big)^2\Big(Z^{(1)}_{[444]}Z^{(0)}_{[444]}\Big)^3
\Big(Z^{(-1)}_{[444]}\Big)^2 Z^{(-2)}_{[444]} \ \ \ \
\nn
\ee
}

\noindent
Like in the previous examples,
clearly seen   is the symmetry between the coefficients in the $A^{2p}$ and $A^{2(|R|-p)}$
terms, typical for binomial-like expansions.
The powers of $q^2$ are just the sums of indices $i$ for all $Z$-factors $Z_{[444]}^{(i)}$ in the products.
Note also additional powers of $Z$-factors, which are not directly predicted by the
group-theory restrictions (\ref{grth444}).

\subsection{List of examples}

It can be convenient to have a collection of the simplest answers brought together.
To preserve maximum of information we give them for the trefoil $3_1$,
in case of $4_1$ one just omits powers of $(-A^2)$ and $q^2$.

\be
H_{[1]}^{3_1}= 1 - A^2 \,[2] \,Z_{[1]}^{(0)}
\nn
\ee

\be
H_{[11]}^{3_1}= 1 - A^2 \,[2] \,Z_{[11]}^{(0)}+q^{-2}A^4  \,Z_{[11]}^{(0)}Z_{[11]}^{(-1)}
\nn
\ee

\be
H_{[22]}^{3_1} = 1  \underline{- [2]^2\, A^2 \,   Z_{[22]}^{(0)}}
+ [3]\,\Big( q^2A^4   Z_{[22]}^{(+1)} + q^{-2}A^4 Z_{[22]}^{(-1)}\Big) \underline{Z_{[22]}^{(0)}}
-\underline{[2]^2}\, A^6 \, \underline{Z_{[22]}^{(+1)}Z_{[22]}^{(0)}Z_{[22]}^{(-1)}}
+ A^8 \, \underline{Z_{[22]}^{(+1)}\big( Z_{[22]}^{(0)}\big)^2 Z_{[22]}^{(-1)}}
\nn
\ee

\be
H_{[33]}^{3_1} = 1 \underline{ - [3][2]\, A^2 \,  Z_{[33]}^{(0)}}
+  \Big([3]^2\,q^2A^4\,   Z_{[33]}^{(+1)} + \frac{[3][4]}{[2]} q^{-2}A^4\, Z_{[33]}^{(-1)}\Big)
\underline{Z_{[33]}^{(0)}}
-  \Big([4]\, q^6A^6 \, Z_{[33]}^{(+2)} + [4][2]^2\,A^6\,Z_{[33]}^{(-1)}   \Big)
\underline{Z_{[33]}^{(+1)}Z_{[33]}^{(0)}}
+ \nn \\
\!\!\!\!\!\!\!\!\!\!\!
+\Big([3]^2\,q^4A^8 \, Z_{[33]}^{(+2)}+ \frac{[3][4]}{[2]}\,A^8\,Z_{[33]}^{(0)}\Big)
\underline{Z_{[33]}^{(+1)}Z_{[33]}^{(0)}Z_{[33]}^{(-1)}}
- [3][2]\,q^4A^{10}\, \underline{Z_{[33]}^{(+2)} Z_{[33]}^{(+1)}}
\big(\underline{Z_{[33]}^{(0)}}\big)^2 \underline{Z_{[33]}^{(-1)}}
+ q^6A^{12}\,
\underline{Z_{[33]}^{(+2)}} \big(\underline{Z_{[33]}^{(+1)} Z_{[33]}^{(0)}}\big)^2
\underline{Z_{[33]}^{(-1)}}
\nn
\ee

%{\footnotesize
\be
H_{[44]}^{3_1} = 1 \underline{ - [4][2]\, A^2 \,  Z_{[44]}^{(0)}}
+  \Big(\frac{[3]^2[4]}{[2]}\,q^2A^4\,   Z_{[44]}^{(+1)} + \frac{[4][5]}{[2]} q^{-2}A^4\, Z_{[44]}^{(-1)}\Big)
\underline{Z_{[44]}^{(0)}}
-\nn \\
-  \Big([4]^2\, q^6A^6 \, Z_{[44]}^{(+2)} + {[5][4][2]}\,A^6\,Z_{[44]}^{(-1)}   \Big)
\underline{Z_{[44]}^{(+1)}Z_{[44]}^{(0)}}
+ \nn \\
+\Big([5]\,q^{12} A^8 \, Z_{[44]}^{(+3)}Z_{[44]}^{(+2)}+ [5][3]^2\,q^4A^8 \, Z_{[44]}^{(+2)}Z_{[44]}^{(-1)}
+ \frac{[5][4]^2}{[2]^2}\,A^8\,Z_{[44]}^{(0)}Z_{[44]}^{(-1)}\Big)
\underline{Z_{[44]}^{(+1)}Z_{[44]}^{(0)}}
- \nn \\
-\Big([4]^2\,q^{10}A^{10}\,Z_{[44]}^{(+3)}+[5][4][2]\,q^4A^{10}\, Z_{[44]}^{(0)}\Big)
\underline{Z_{[44]}^{(+2)} Z_{[44]}^{(+1)}Z_{[44]}^{(0)} Z_{[44]}^{(-1)}}
+ \nn \\
+ \Big(\frac{[3]^2[4]}{[2]}\,q^{10}A^{12}\,Z_{[44]}^{(+3)} + \frac{[5][4]}{[2]}\,q^6A^{12}\,Z_{[44]}^{(+1)}\Big)
\underline{Z_{[44]}^{(+2)} Z_{[44]}^{(+1)}}\big(\underline{Z_{[44]}^{(0)}}\big)^2 \underline{Z_{[44]}^{(-1)}}
- \nn \\
- [4][2]\,q^{12}A^{14}
\underline{Z_{[44]}^{(+3)}Z_{[44]}^{(+2)}}\big(\underline{ Z_{[44]}^{(+1)}Z_{[44]}^{(0)}}\big)^2
\underline{Z_{[44]}^{(-1)}}
+ q^{16}A^{16}\,
Z_{[44]}^{(+3)}\big(\underline{Z_{[44]}^{(+2)} Z_{[44]}^{(+1)} Z_{[44]}^{(0)}}\big)^2
\underline{Z_{[44]}^{(-1)}}
\nn
\ee
%}

\be
H_{[333]}^{3_1} = 1 -A^2\, [3]^2\, Z_{[333]}^{(0)}
+  A^4\,\frac{[4][3]^2}{[2]}\Big(q^2Z^{(1)}_{[333]} + q^{-2}Z^{(-1)}_{[333]}\Big)\,Z^{(0)}_{[333]}
-\nn\\
- A^6 \left(\frac{[5][4]}{[2]}
\Big(q^6 Z^{(2)}_{[333]}Z^{(1)}_{[333]}+q^{-6}Z^{(-2)}_{[333]}Z^{(-1)}_{[333]}\Big)Z^{(0)}_{[333]}+
[4]^2[2]^2Z^{(1)}_{[333]}Z^{(0)}_{[333]}Z^{(-1)}_{[333]}\right)
+ \nn \\
+ A^8\left( [3]^2[5]\Big(q^4Z^{(2)}_{[333]}+q^{-4}Z^{(-2)}_{[333]}\Big)
+ \frac{[3]^2[4]^2}{[2]^2}  Z^{(0)}_{[333]}  \right)Z^{(1)}_{[333]}Z^{(0)}_{[333]}Z^{(-1)}_{[333]}
- \nn \\
-A^{10}\left( [3]^2[5]
\Big(q^{4}Z^{(2)}_{[333]}+q^{-4}Z^{(-2)}_{[333]} \Big)Z^{(0)}_{[333]}
+ \frac{[3]^2[4]^2}{[2]^2} Z^{(2)}_{[333]}Z^{(-2)}_{[333]}  \right)Z^{(1)}_{[333]}Z^{(0)}_{[333]}Z^{(-1)}_{[333]}
+ \nn \\
%\!\!\!\!\!\!\!\!\!\!\!\!\!\!\!\!\!\!\!\!\!\!\!\!\!\!\!\!\!\!
+A^{12}\left(\frac{[5][4]}{[2]}
\Big(q^{6} Z^{(2)}_{[333]}Z^{(1)}_{[333]}
+q^{-6}Z^{(-2)}_{[333]}Z^{(-1)}_{[333]} \Big) +
[4]^2[2]^2Z^{(2)}_{[333]}   Z^{(-2)}_{[333]}\right)
Z^{(1)}_{[333]}\Big(Z^{(0)}_{[333]}\Big)^2Z^{(-1)}_{[333]}
- \nn \\
-A^{14}\,\frac{[3]^2[4]}{[2]}\Big(q^{2} Z^{(1)}_{[333]}   +
q^{-1} Z^{(-1)}_{[333]}   \Big)\,
Z^{(2)}_{[333]}Z^{(1)}_{[333]}\Big(Z^{(0)}_{[333]}\Big)^2Z^{(-1)}_{[333]}Z^{(-2)}_{[333]}
+ \nn\\
+\Big(A^{16}[3]^2 -A^{18}Z^{(0)}_{[333]}\Big)
Z^{(2)}_{[333]}\Big(Z^{(1)}_{[333]}Z^{(0)}_{[333]}Z^{(-1)}_{[333]}\Big)^2Z^{(-2)}_{[333]}
\nn
\ee

\section{Conjecture for generic rectangular diagram $R=[r^s]$}

\subsection{The structure of $Z$-factors}

As already stated, we assume that for defect-zero knots ${\cal K}^{(0)}$,
i.e. those with the fundamental Alexander of degree one,
$Al_{_\Box}^{{\cal K}^{(0)}}  = \alpha + \beta(q^2+q^{-2}) $,
the rectangular colored HOMFLY  depend only on the shifted $Z$-factors
$Z_{[r^s]}^{(i)} = \{Aq^{r+i}\}\{Aq^{i-s}\}$
(among other things this implies \cite{IMMMfe} a simple conjecture for the superpolynomials,
because $Z$-factors, in variance of individual differentials,
are easily made positive after $T$-deformation).
The first question is {\it which} of these $Z$-factors actually contribute.

\bigskip

Group theory restrictions in the case of $R=[r^s]$ predict the occurrence of factors
up to $\{A/q^{2s-1}\}$ from $[r^s] \ \stackrel{SL(2s-1)}{\cong} \ [r^{s-1}]$
and up to $\{Aq^{2r-1}\}$ from
$\widetilde{[r^s]} = [s^r] \ \stackrel{SL(2r-1)}{\cong} \ [s^{r-1}] = \widetilde{[r-1]^s}$.
Since these factors belong respectively to $Z_{r^s]}^{(1-s)} = \{Aq^{r+1-s}\}\{A/q^{2s-1}\}$
and $Z_{[r^s]}^{(r-1)} = \{Aq^{2r-1}\}\{Aq^{r-1-s}\}$, all shifts
$i$ between $1-s$ and $r-1$
unavoidably appear in the expansion of $H_{[r^s]}^{{\cal K}^{(0)}}$.
We conjecture that nothing more actually shows up.

\bigskip

\noindent
A more extended conjecture includes the following theses:

$\bullet$ $H_{[r^s]}^{{\cal K}^{(0)}}$ is a polylinear combination of shifted factors
$Z^{(i)}_{[r^s]} = \{Aq^{r+i}\}\{A/q^{s-i}\}$

$\bullet$ The contributions of the order $A^{2p}$ is a linear combination
of items, each containing a product of $p$ such $Z$-factors

$\bullet$ Each item is proportional to a product of several chains $\prod_{i=i_L}^{i_R} Z^{(i)}_{[r^s]}$

$\bullet$ Chain has no gaps, it obligatory includes $i=0$ and is restricted
as stated above, i.e. $-(s-1)\geq i_L\leq 0$ and $0\leq i_R\leq r-1$

$\bullet$ The chains form "floors", and each floor is shorter at least by two,
so that there are no two-step edges in the pyramid, see (\ref{pictorH}) below

$\bullet$ The number of floors can not exceed ${\rm min}(r,s)$

\bigskip

\noindent
From these rules it follows that
\be
H_{[r]}^{{\cal K}^{(0)}} = \oplus_{j=0}^{r} \prod_{i=0}^{j-1} Z_{[r]}^{(i)}
= \oplus_{j=0}^{r} \prod_{i=0}^{j-1} \{Aq^{r+i}\}\{Aq^{i-1}\} \nn \\
H_{[1^s]}^{{\cal K}^{(0)}} = \oplus_{j=0}^s \prod_{i=0}^{j-1} Z_{[1^s]}^{(-i)}
= \oplus_{j=0}^s \prod_{i=0}^{j-1} \{Aq^{1-i}\}\{A/q^{s+i}\}
\nn \\ \nn \\
H_{[rr]}^{{\cal K}^{(0)}} = \ \
\oplus_{ -1\leq i_L\leq 0\leq i_R\leq r-1} \prod_{i=i_L}^{i_R} Z_{[rr]}^{(i)}
\ \ \ \ \
\oplus_{ -1<i_{L1}<i_{L2}\leq  0\leq i_{R2}<i_{R1}\leq r-1}
\prod_{i_1=i_{L1}}^{i_{R1}} Z_{[rr]}^{(i_1)}
\prod_{i_2=i_{L2}}^{i_{R2}} Z_{[rr]}^{(i_2)}
\nn \\ \nn \\
\ldots
\ee
what is indeed true in numerous tested examples.

In pictorial form
\be
H_{[r^s]}^{{\cal K}^{(0)}} =
1 -A^2\, \boxed{0}  + A^4 \Big(\boxed{10} \ \oplus \ \boxed{0{-1}}\Big)
-A^6 \Big(\boxed{210} \ \oplus \ \boxed{10{-1}} \ \oplus \ \boxed{0{-1}{-2}}\Big) + \nn \\ \nn \\
+A^8\left(\boxed{3210} \ \oplus \ \boxed{210{-1}} \ \oplus {\boxed{0}\over\boxed{10\,{-1}}} \
\oplus \ \boxed{10{-1}{-2}} \ \oplus \ \boxed{0{-1}{-2}{-3}}\right) - \nn \\ \nn\\
- A^{10}\left(\boxed{43210} \ \oplus\ \boxed{3210\,{-1}}
\ \oplus{\boxed{0}\over\boxed{210\,{-1}}} \ \oplus \ \boxed{210\,{-1}\,{-2}}
 \ \oplus \ {\boxed{0}\over\boxed{10\,{-1}\,{-2}}}
\ \oplus\ \boxed{10\,{-1}\,{-2}\,{-3}} \ \oplus\ \boxed{0\,{-1}\,{-2}\,{-3}\,{-4}}
\right) +
\nn
\ee
\be
+\ \ldots \ -\  A^{18}\left(
\boxed{876543210}\ \oplus \
\ldots \ \oplus \  { {\boxed{0}\over\boxed{10\,{-1}}} \over \boxed{210\,{-1}\,{-2}} }
\ \oplus\ \ldots \ \oplus\ \boxed{0\,{-1}\,{-2}\,{-3}\,{-4}\,{-5}\,{-6}\,{-7}\,{-8}} \right) + \ \ldots
\label{pictorH}
\ee
where shown in the boxes are the  shifts $\{i\}$, each item with a set $I$ inside the boxes
stands for the product
$\prod_{i\in I} Z^{(i)}_{[r^s]}$ with some yet unspecified $q$-dependent coefficients.
This pictorial expansion does not depend on $r$ and $s$ -- but actually
contributing are only the items with all entries $i$ within the range $-s<i<r$.
Clearly, it follows that the number of floors does not exceed  ${\rm min}(r,s)$.

\subsection{Coefficients}

We now need to substitute direct-sum symbols in (\ref{pictorH}) by concrete $q$-dependent
coefficients, which depend on $r$ and $s$, in particular, explicitly respect the selection rules,
conjectured in the previous subsection.

The structure of the formula is already clear from above examples:

{\footnotesize
\be
H_{[r^s]}^{{\cal K}^{(0)}} =
1 -A^2\, [r][s]\, \boxed{0}
+ A^4 \left(\frac{[r][r-1]}{[2]}\frac{[s][s+1]}{[2]}\,\boxed{10} \ + \
\frac{[r][r+1]}{[2]}\frac{[s][s-1]}{[2]}\,\boxed{0{-1}}\right) - \nn \\ \nn \\ \nn \\
-A^6 \left(\frac{[r][r-1][r-2]}{[2][3]}\frac{[s][s+1][s+2]}{[2][3]}\,\boxed{210} \ + \
[2]^2\frac{[r+1][r][r-1]}{[2][3]}\frac{[s+1][s][s-1]}{[2][3]}\,\boxed{10{-1}} \ + \
\frac{[r][r+1][r+2]}{[2][3]}\frac{[s][s-1][s-2]}{[2][3]}\,\boxed{0{-1}{-2}}\right) + \nn \\ \nn \\ \nn \\
+A^8\left(\frac{[r][r-1][r-2][r-3]}{[2][3][4]}\frac{[s][s+1][s+2][s+3]}{[2][3][4]}\,\boxed{3210} \ + \
[3]^2\frac{[r+1][r][r-1][r-2]}{[2][3][4]}\frac{[s+2][s+1][s][s-1] }{[2][3][4]}\,\boxed{210{-1}} \ +
\right. \nn \\  +
\frac{[r][s]}{[2]^2}\frac{ [r+1][r][r-1]}{[2][3] }\frac{[s+1][s][s-1]  }{[2][3] }
\, {\boxed{0}\over\boxed{10\,{-1}}} \ +
\nn \\ \left. +
[3]^2\frac{[r+2][r+1][r][r-1]}{[2][3][4]}\frac{[s+1][s][s-1][s-2] }{[2][3][4]}\,\boxed{10{-1}{-2}} \ +
\frac{[r][r+1][r+2][r+3]}{[2][3][4]}\frac{[s][s-1][s-2][s-3]}{[2][3][4]}\,\boxed{0{-1}{-2}{-3}}\right)
- \nn \\ \nn\\ \nn \\
- A^{10}\left(C^5_r C^5_{s+4}\, \boxed{43210} \ + \ [4]^2 C^5_{r+1} C^5_{s+3}\, \boxed{3210\,{-1}}
\ + \ C^1_rC^1_sC^4_{r+1}C^4_{s+2} {\boxed{0}\over\boxed{210\,{-1}}} \ + \ [6]^2 C^5_{r+2}C^5_{s+2}\, \boxed{210\,{-1}\,{-2}}
+ \right. \nn \\ \left.
 \ + \ C^1_rC^1_sC^4_{r+2}C^4_{s+1} {\boxed{0}\over\boxed{10\,{-1}\,{-2}}}
\ +\ [4]^2 C^5_{r+3}C^5_{s+1}\, \boxed{10\,{-1}\,{-2}\,{-3}} \ +\
C^5_{r+4}C^5_s\,\boxed{0\,{-1}\,{-2}\,{-3}\,{-4}}
\right) +
\nn
\ee
\be
+\ \ldots \ -
\nn\ee
\be
-\  A^{18}\left(
C^9_r C^9_{s+8}\, \boxed{876543210}\ + \
\ldots \ + \   \frac{[2]^2[6]^2}{[3]^4[4]^4}C^1_rC^1_sC^3_{r+1}C^3_{s+1}C^5_{r+2}C^5_{s+2}
 { {\boxed{0}\over\boxed{10\,{-1}}} \over \boxed{210\,{-1}\,{-2}} }
\ +\ \ldots \ +\ C^9_{r+8}C^9_s\,\boxed{0\,{-1}\,{-2}\,{-3}\,{-4}\,{-5}\,{-6}\,{-7}\,{-8}} \right) + \ \ldots
\nn
\ee
}
\noindent
where the quantum binomial coefficients $C^m_n = \frac{[n]!}{[m]![n-m]!}$ are defined to vanish for $m>n$,
like  for $q=1$.

%It is clear that all the single line (single floor) contributions come with the coefficients ???

The general term  of the expansion can be parameterized as follows:
\be
H^{3_1}_{[r^s]} =
\!\sum_{\{a,b\}} W\{a,b\}\cdot
%\left(
\!\!\!\!\!\prod_{f=1}^{\!{\rm min}(r,s)}
\boxed{(-A^2)^{p_f} q^{p_f(a_f- b_f)}}\cdot
\Big(C_{a_f+b_f}^{b_f}\Big)^2 C_{r+b_f}^{p_f} C_{s+a_f}^{p_f}
%\right)
\cdot
{\tiny
{{{{{
\ldots\phantom{\oint^{\oint^5_5}_{\oint^5_5}}
\over \boxed{a_f\ldots 0 \ldots -b_f}   }
\over  \ldots\phantom{\oint^{\oint^5_5}_{\oint^5_5}}  }
\over\boxed{a_3\ \ \ \ \ldots \ \ \ \ 1\, 0\,{-1}\ \ \ldots \ \ \ \ -b_3}  }
\over\boxed{a_2\ \ \ \ \ \ \ldots  \ \ \ \ 1\, 0\,{-1}\ \  \ \ \ldots\ \ \  \ \ -b_2}  }
\over\boxed{a_1\ \ \ \ \ \ \   \ \ldots \ \ \ \ \ \ \ 1\, 0\,{-1}\ \ \ \ \ \ \ \ldots\ \ \ \ \ \ \ \ -b_1}}
}
\label{H31rspict}
\ee
with the number of $Z$-factors at the $f$-th floor equal to $p_f=a_f+b_f+1$.
Actually, the product of binomial coefficients is
\be
\prod_f \Big(C_{a_f+b_f}^{b_f}\Big)^2 C_{r+b_f}^{p_f} C_{s+a_f}^{p_f}
= \prod_f  \frac{1}{([a_f]![b_f]! [a_f+b_f+1])^2}
\frac{[r+b_f]![s+a_f]!}{[r-1-a_f]![s-1-b_f]!}
%= \left(\frac{[a_f+b_f]!}{[a_f]![b_f]!}\right)^2 C_{r+b_f}^{p_f} C_{s+a_f}^{p_f}
\ee
while the picture stands for the product
\be
\prod_f \left(\prod_{i_f=-b_f}^{a_f} Z_{r^s}^{(i_f)}\right) =
\prod_f \left(\prod_{i_f=-b_f}^{a_f} \{Aq^{r+i_f}\}\{Aq^{i_f-s}\}\right)
\ee
These contributions should be now summed over all $a_f$ and $b_f$,
constrained by the pyramid conditions
$\ \ldots < a_f <\ldots < a_3<a_2<a_1<r\ $ and  $\ \ldots < b_f < \ldots < b_3< b_2<b_1<s$.
These constraints are automatically imposed by the weights $W\{a,b\}$ in the sum.

For a single-floor contributions there are no non-trivial weights, $W=1$.

For two floors
\be
W\{a_1,b_1|a_2,b_2\} = \left(\frac{[a_1-a_2][b_1-b_2]}{[a_1+b_2+1][a_2+b_1+1]}\right)^2
\ee
For three floors the weight factor is a product of two-floor factors:
\be
W\{a_1,b_1|a_2,b_2|a_3,b_3\} = W\{a_1,b_1|a_2,b_2\} W\{a_2,b_2|a_3,b_3\} W\{a_1,b_1| a_3,b_3\}
%\nn \\
%\xi\{2,2|1,1|0,0\} = \frac{[2]^4}{[3]^4}, \ \ \ \ \ \
%W\{2,2|1,1|0,0\} = W\{2,2|1,1\}W\{1,1|0,0\}\xi\{2,2|1,1|0,0\} =
%\frac{1}{[4]^4}\frac{1}{[2]^4}\frac{[2]^4}{[3]^4} = \frac{1}{[3]^4[4]^4}
%???
 \ee
and in general
\be
W\{a,b\} = \prod_{f'<f''} W\{a_{f'},b_{f'}|a_{f''},b_{f''}\}
\ee
Putting everything together we obtain for the differential expansion in the case of trefoil:

\be
H^{3_1}_{[r^s]}= \sum_{F=1}^{{\rm min}(r,s)} \sum_{
\stackrel{0\leq a_F <\ldots < a_3<a_2<a_1<r}{0\leq b_F < \ldots < b_3< b_2<b_1<s}}
\ \ \prod_{f'<f''}^F W\{a_{f'},b_{f'}|a_{f''},b_{f''}\}\cdot
\label{de31rs}
\ee
\vspace{-0.2cm}
\be
\cdot \prod_{f=1}^F \left(\boxed{(-A^2)^{p_f} q^{p_f(a_f- b_f)}}\cdot
\left(\frac{[a_f+b_f]!}{([a_f]![b_f]! }\right)^2
\frac{[r+b_f]![s+a_f]!}{[r-1-a_f]![s-1-b_f]!\big([a_f+b_f+1]!\big)^2}
 \prod_{i_f=-b_f}^{a_f} \{Aq^{r+i_f}\}\{Aq^{i_f-s}\}\right)
 \nn
\ee
\vspace{0.4cm}
We tested this formula up to $R=[8^3]$, $R=[6^4]$ and $R=[5^5]$.

\bigskip

The answer for $H^{4_1}_{[r^s]}$ is conjecturally obtained by elimination of the boxed factor
from (\ref{H31rspict}):
\be
\boxed{
H^{4_1}_{[r^s]} \stackrel{?}{=}
\sum_{\{a,b\}} W\{a,b\}\cdot
\left(\prod_{f=1}^{{\rm min}(r,s)}
%\boxed{(-A^2)^{p_f} q^{p_f(a_f- b_f)}}\cdot
\Big(C_{a_f+b_f}^{b_f}\Big)^2 C_{r+b_f}^{p_f} C_{s+a_f}^{p_f}
\right)\cdot
{\footnotesize
{{{{{
\ldots\phantom{\oint^{\oint^5_5}_{\oint^5_5}}
\over \boxed{a_f\ldots 0 \ldots -b_f}   }
\over  \ldots\phantom{\oint^{\oint^5_5}_{\oint^5_5}}  }
\over\boxed{a_3\ \ \ \ \ldots \ \ \ \ 1\, 0\,{-1}\ \ \ldots \ \ \ \ -b_3}  }
\over\boxed{a_2\ \ \ \ \ \ \ldots  \ \ \ \ 1\, 0\,{-1}\ \  \ \ \ldots\ \ \  \ \ -b_2}  }
\over\boxed{a_1\ \ \ \ \ \ \   \ \ldots \ \ \ \ \ \ \ 1\, 0\,{-1}\ \ \ \ \ \ \ \ldots\ \ \ \ \ \ \ \ -b_1}}
}
}
%\label{H41rspict}
\nn
\ee
\be
= \sum_{F=1}^{{\rm min}(r,s)} \sum_{
\stackrel{0\leq a_F <\ldots < a_3<a_2<a_1<r}{0\leq b_F < \ldots < b_3< b_2<b_1<s}}
\ \ \prod_{f'<f''}^F W\{a_{f'},b_{f'}|a_{f''},b_{f''}\}\cdot
\label{41rs}
\ee
\vspace{-0.2cm}
\be
\cdot \prod_{f=1}^F \left(
\left(\frac{[a_f+b_f]!}{([a_f]![b_f]! }\right)^2
\frac{[r+b_f]![s+a_f]!}{[r-1-a_f]![s-1-b_f]!\big([a_f+b_f+1]!\big)^2}
 \prod_{i_f=-b_f}^{a_f} \{Aq^{r+i_f}\}\{Aq^{i_f-s}\}\right)
 \nn
\ee

\bigskip

\noindent
Particular polynomials, calculated with the help of this formula,
satisfy the standard tests, described in sec.6 of \cite{mmms1}.

\section{Conclusion
\label{conc}}

In this paper we made a very plausible conjecture for explicit
formulas for rectangularly-colored HOMFLY polynomials for the figure-eight
knot $4_1$.
Further conjecture for the corresponding superpolynomials and 4-graded
hyperpolynomials of \cite{GGS} should follow, according to  \cite{IMMMfe} and \cite{arthdiff}.

Conjecture is made on the basis of study of differential expansions,
which are especially simple for defect-zero knots
and, moreover, are nearly identical for $4_1$ and for the trefoil $3_1$.
Arbitrarily-colored HOMFLY are known for trefoil (as well as for any other torus knots)
from the Rosso-Jones formula \cite{RJ,DMMSS},
thus the only non-trivial exercise is to convert it into a differential expansion form.
This is indeed quite a tedious job, and it is described in the present paper.
The result is eq.(\ref{41rr}) for the $R=[rr]$ and eq.(\ref{41rs}) for generic rectangular
$R=[r^s]$.
It directly generalizes the archetypic expression of \cite{IMMMfe} for symmetric $R=[r]$
and antisymmetric $R=[1^r]$ representations.

Further generalizations are needed in three directions:

$\bullet$ to non-rectangular diagrams

$\bullet$ to other knots with defect  \cite{Konodef} zero

$\bullet$ to all knots

Each of these directions faces immediate difficulties.
Hopefully, they will be resolved in the near future.

\section*{Acknowledgements}

I am indebted to A.Sleptsov for help with the calculation of colored HOMFLY for the trefoil
from the Rosso-Jones formula and also to A.Mironov, S.Nawata and P.Ramadevi for encouraging
questions.

This work was performed at the Institute for the Information Transmission Problems
with the  support from the Russian Science Foundation, Grant No.14-50-00150.

\end{document}